\documentclass[a4paper,fleqn,usenatbib]{mnras}

\usepackage{newtxtext,newtxmath}

\usepackage[T1]{fontenc}
\usepackage{ae,aecompl}

\usepackage{graphicx}	
\usepackage{amsmath}	
\usepackage{amssymb}	
\usepackage{bm}

\newcommand{\pyr}{\,yr$^{-1}$}  
\newcommand{\msun}{\,M$_{\sun}$}
\newcommand{\ps}{\,s$^{-1}$}
\newcommand{\cmspg}{\,cm$^{2}$\,g$^{-1}$}
\newcommand{\gcc}{\,g\,cm$^{-3}$}

\title[Mass Tran. and Disc Formation in AGB Binary Sys.]{Mass Transfer and Disc Formation in AGB Binary Systems}

\author[Zhuo Chen et al.]{
Zhuo Chen,$^{1}$\thanks{E-mail: zchen25@ur.rochester.edu}
Adam Frank,$^{1}$\thanks{E-mail: afrank@pas.rochester.edu}
Eric G. Blackman$^{1}$
Jason Nordhaus$^{2, 3}$\newauthor
and Jonathan Carroll-Nellenback$^{1}$
\\
$^{1}$Department of Physics and Astronomy, University of Rochester, NY 14627, USA\\
$^2$National Technical Institute for the Deaf, Rochester Institute of Technology, NY 14623, USA\\
$^3$Center for Computational Relativity and Gravitation, Rochester Institute of Technology, NY 14623, USA\\
}

\date{Accepted XXX. Received YYY; in original form ZZZ}

\pubyear{2017}

\begin{document}
\label{firstpage}
\pagerange{\pageref{firstpage}--\pageref{lastpage}}
\maketitle

\begin{abstract}
We investigate  mass transfer and the formation of disc in binary systems using a combination of numerical simulations  and theory. We consider six models distinguished by  binary separation, secondary mass and outflow mechanisms. Each system consists of an asymptotic-giant-branch (AGB) star and an accreting secondary. The AGB star loses its mass via a wind. In one of our six models, the AGB star incurs a short period of outburst. In all cases, the secondary accretes part of the ejected mass and also influences the mass-loss rate of the AGB star. The ejected mass may remain gravitationally bound to the binary system and form a circumbinary disk, or contribute to an accretion disk around the secondary. In other cases, the ejecta will escape the binary system. The accretion rate on to the secondary changes non-linearly with  binary separation. In our closest binary simulations, our models exemplify the wind Roche lobe overflow while in our wide binary cases, the mass transfer exhibits Bondi-Hoyle accretion. The morphologies of the outflows in the binary systems are varied. The variety may provide clues to how  the late AGB phase influences planetary nebulae shaping. We employ the adaptive-mesh-refinement code {\small ASTROBEAR} for our simulations and include ray-tracing, radiation transfer, cooling and dust formation. To attain the highest computational efficiency and the most stable results, all simulations are run in the corotating frame.
\end{abstract}




\section{Introduction}

More than 50 per cent of the total population of  stars are found in binary systems. During most of their main sequence lifetime, they do not exchange mass but when one of the star progresses on to the giant branch [the red giant branch (RGB) or later asymptotic giant branch (AGB) phases], its radius grows large enough that it may interact with the secondary. Mass transfer can ensue.

Single AGB stars will lose mass by driving AGB winds modulated  with pulsations \citep{habing2013}. Typical mass loss rates and wind speeds are $10^{-7}-10^{-5}$\msun\pyr and  $10-20$ km\ps, respectively \citep{herwig2005}. The standard  theory of AGB wind formation can be described as follows: the pulsation in the AGB star's atmosphere  drives shocks to large radii.   The post-shock, compressed  gas will  form dust grains when cooling drops the equilibrium temperature  below the dust condensation temperature. In turn, the dust grains will absorb radiation from the AGB star and accelerate.  As the accelerated dust grains  carry the gas with them,  the AGB wind is  formed. This theory has found considerable phenomenological success \citep{bowen1988,liljegren2016}. A lingering question is where does the wind pulsation come from? The answer  likely involves convection \citep{freytag2008,arroyotorres2015} and radiation transfer. Convection is a   complex problem and in the present absence of a detailed convective AGB stellar atmosphere model  we adopt a phenomenological AGB pulsation wind model in our simulations.

Gas  ejected by the AGB wind in  binary can be accreted by the secondary. Gas that is not accreted can remain gravitationally bound to the binary system or eventually escape. There are three important length scales in this context: the radius of the giant star, the  dust formation radius, and the binary separation. Different combinations of these length scales lead to fundamentally different  mass transfer mechanisms. If the binary separation is much larger than the dust formation radius,  the mechanisms of accretion on to the secondary will be BH accretion. If the binary separation is slightly larger than the dust formation radius, the accretion process will be wind Roche lobe overflow (WRLOF). And if the binary separation is less than the dust formation radius, Roche lobe overflow (RLOF) may incur. When both stars' Roche-lobes are filled with gas, a common envelope (CE) phase will commence. 

BH accretion is perhaps the most thoroughly studied of the aforementioned  accretion mechanisms \citep{edgar2004}. The classic BH problem \citep{bondi1944} has an accreting object in parallel flow and predicts that the gas within a cylinder will be focused towards the back of the object. The flow rate within that cylinder is the upper limit of the accretion rate. This model is suitable for an  isolated star travelling through interstellar medium or X-ray binaries \citep{pfahl2002}. In intermediate separation binary systems, gas flow (wind) become non-parallel and the full picture of binary systems is needed. \citet{huarte2013} considered the wind trajectory and simulated the 3D wind-capture disc in binary systems. However, they solved the problem in a corotating box just around the secondary. In this paper, we focus on the holistic processes involving both the mass-losing star and the accreting star. The dynamics of the wind will be self-consistently solved.

\citet{podsiadlowski2007} put forward the concept of WRLOF. They argued that WRLOF could be a very efficient way of mass transfer mechanism in binaries. In a more general picture, \citet{dermine2009} concluded that the Roche lobe can be substantially modified by radiation force. They also realized that the Roche lobe model may not be useful at all in very luminous systems (e.g. AGB stars). Despite the possible intricacy, WRLOF model was applied to study the population of carbon-enhanced metal-poor stars and got positive results \citep{abate2013}. However, a more precise and global picture of WRLOF is still needed. Specifically, we treat the binary stars and the AGB wind as an integrated system and study their mutual influence. That is the interaction we want to address.

RLOF only happens when the Roche lobe surface is connected and WRLOF happens when the Roche lobe surface is detached. \citet{schuerman1972} illustrated the  boundary of the two scenarios with a critical value $\delta_\text{c}$. When the ratio of radiation force to gravity is larger than $\delta_\text{c}$, the Roche lobe surface is detached \citep{dermine2009}. To estimate the size of the Roche lobe, \citet{eggleton1983} approximated the radii of the Roche lobe with a smooth function that became useful in many subsequent studies \citep{staff2016a}. RLOF model was also applied to X-ray binaries by \citet{savonije1978} and was able to match the observation. A pioneering hydrodynamical study on RLOF was performed by \citet{sawada1986}. Their 2D hydrodynamical simulations revealed the complex flow pattern in a close binary system.

The concept of CE evolution was formally proposed by \citet{paczynski1976}. It is a short lived and rapidly evolving (e.g. rapidly decreasing orbital separation, ejection of the envelope or tidal destruction) phase \citep{ivanova2013} in binary evolution.The progenitors of CE and planetary nebulae (PNe) are possibly very similar and a link has been confirmed observationally \citep{hillwig2016}. However, the CE phase is still poorly understood and attracting increasing attention \citep{passy2012,demarco2016}. \citet{nordhaus2006}, \citet{ivanova2016} and \citet{nandez2016} investigated the dynamics of in-spiral of the companion and the ejection of the envelop during the in-fall. \citet{staff2016b} performed a 3D hydrodynamic simulation  and validated the in-fall process of a $10M_\text{J}$ planet into the envelope of an RGB star and an AGB star.

As a consequence of the aforementioned four mass transfer mechanisms, binary systems are the progenitors of many other objects.

PNe are one class of successor of binary systems. Various PNe shapes and shaping mechanism were discussed\citep{balick2002}. It is believed that disc and tori are common in the progenitors of PNe \citep{winckel2003,ruyter2006,winckel2009,bujarrabal2013} and they are important in the evolution of PNe. A pioneering work in validating this assumption was done by \citet{mastrodemos1998} and continuous efforts have been devoted to study the formation of circumbinary discs in AGB binary systems \citep{akashi2008,akashi2015,chen2016a}. Besides PNe, Type Ia supernovae are also realized to be important successors of binary systems \citep{whelan1973,iben1984}. Researches on Type Ia supernovae focus on the explosion more than on the initial condition and a satisfactory answer can only be found with realistic binary models.

For our binary calculations, we use  {\small ASTROBEAR} \footnote{https://astrobear.pas.rochester.edu/} \citep{cunningham2009,carroll2013}. {\small ASTROBEAR} is a magneto-hydrodynamic code that has adaptive-mesh-refinement (AMR) capability. AMR allows us to only refine the region that has the highest variable (density, pressure etc.) gradients. Every additional level of AMR means the doubling of resolution of every dimension. Thus the effective resolution of a 3D simulation with n levels of AMR is $8^n$ times higher than the same one without AMR but the with the same base grid. This technique can speed up the binary calculation by more than $10^5$ times compared to a uniform grid code. We perform our calculations on University of Rochester's BlueHive supercomputer cluster \footnote{https://info.circ.rochester.edu/}. Each binary model (without a mass loss 'burst') takes roughly 240 h on 240 cores and the binary model with a burst takes 300 h on 240 cores. The equivalent total computational time for the six simulations is $4.18\times10^{5}$ CPU hours.

In this paper, we present the first global 3D radiation hydrodynamic calculation of mass transfer in AGB binary systems. Section \ref{sec:model} contains the equations and physics in our calculation. A description of the isolated AGB star is also included in section~\ref{sec:model}. Section~\ref{sec:binary} will present the binary simulation results: the affected mass loss rate , the accretion rate and the morphology of the outflow. We also qualitatively analyse and classify the simulations in Section~\ref{sec:binary}.  Section~\ref{sec:summary} has the conclusions and discussion.

\section{model description}\label{sec:model}

We now present the equations used in our calculation and explain the underlying physics and approximations. In what follows, we describe both our isolated AGB wind model and the implementation of the 'burst'.

\subsection{Equations and physics}

\subsubsection{Equations}

We use radiation hydrodynamic equations of an ideal gas:

\begin{eqnarray}
    \frac{\upartial\rho}{\upartial t}+\nabla\cdot\left(\rho\bm{u}\right)&=&0\label{eqn:mass}\\ 
    \frac{\upartial\rho\bm{u}}{\upartial t}+\nabla\cdot\left(\rho\bm{uu}\right)&=&-\nabla p+\rho\left(\bm{f}_\text{g}+\bm{f}_\text{rad}+\bm{f}_\text{ni}\right)\label{eqn:mom}\\ 
    \frac{\upartial E}{\upartial t}+\nabla\cdot\left[\left(E+p\right)\bm{u}\right]&=&\rho\left[\left(\bm{f}_\text{g}+\bm{f}_\text{rad}+\bm{f}_\text{ni}\right)\cdot\bm{u}-\dot{Q}\right]\label{eqn:energy}\\ 
    p&=&nk_\text{b}T
\end{eqnarray}
In above equations, $\rho,n,\bm{u},p,T,E,\dot{Q}$ (see section \ref{sec:cooling}) are the density, particle number density, velocity, pressure, temperature, total energy and cooling rate in the corotating frame respectively. $k_\text{b}$ is the Boltzmann constant. The subscript ni stands for non-inertial terms due to carrying out the calculation in the rotating frame.
\begin{eqnarray}
    n&=&\frac{\rho}{\mu m_\text{H}}\\
    \bm{f}_\text{g}&=&\alpha\frac{G M_\text{AGB}}{r_\text{AGB}^{2}}\hat{\bm{r}}_\text{AGB}+\frac{G M_\text{sec}}{r_\text{sec}^{2}}\hat{\bm{r}}_\text{sec}\\
    \bm{f}_\text{rad}&=&\frac{L\kappa\exp(-\tau)}{4\uppi c r^{2}_\text{AGB}}\hat{\bm{r}}_\text{AGB}\label{eqn:radforce}\\
    \bm{f}_\text{ni}&=&-\bm{\Omega}\times\left(\bm{\Omega}\times\bm{r}\right)-2\bm{\Omega}\times\bm{u}\\
    \bm{\Omega}&=&\left(0,0,\sqrt{\frac{G(M_\text{AGB}+M_\text{sec})}{d^3}}\right)^{T}\\
    E&=&e+\frac{\rho u^2}{2}\\
    e&=&\frac{p}{\rho\left(\gamma-1\right)}
\end{eqnarray}
$m_{H}=1.673\times10^{-24}$g is the mass of hydrogen atom and $\mu=1.3$ is the mean atomic weight. $d$ is the initial binary separation. The origin is set at binary system's centre of mass. The unit vectors in their explicit forms are:
\begin{eqnarray}
    \hat{\bm{r}}_\text{AGB}&=&\frac{\bm{r}_\text{AGB}-\bm{r}}{\left|\bm{r}_\text{AGB}-\bm{r}\right|}\\
    \hat{\bm{r}}_\text{sec}&=&\frac{\bm{r}_\text{sec}-\bm{r}}{\left|\bm{r}_\text{sec}-\bm{r}\right|}
\end{eqnarray}

The quantity $\bm{f}_\text{g}$ is the binary stars' force on the gas; $\bm{f}_\text{rad}$ is the radiation force on the gas; $M_\text{AGB}, M_\text{sec}$ are the masses of the AGB star and the secondary star, respectively; $\alpha$ is a coefficient which will be explained in section \ref{sec:isoagb}; $\kappa$ (see section \ref{sec:dust}) is the Planck mean opacity that changes when dust forms; $L$ is the AGB star's luminosity; $c$ is the speed of light; $\tau$ is the optical depth (see section \ref{sec:rad}); $e$ is the internal energy of the gas; and $\gamma=\frac{5}{3}$ is the ratio of specific heat.

\subsubsection{Gravity}

We solve the gravity interaction using Newton's laws. 
We include the gravitational influence of the stars on the gas and the stars on each other but the stars are not influenced by  gravity of the gas in our model. The calculation would be much less efficient and the model would be more complicated were we to  take the gas self-gravity and its influence on the stars into account. The gravitational feedback from gas to stars is, however, unimportant on time-scales of our simulation ($t\sim100$yr).  It may be important over longer times such as that of the stellar  lifetime. Gravitational interaction between gas and stars can, for example, contribute to the binaries' orbital decay. 
 We will discuss the binaries' orbital change problem in a subsequent paper.

\subsubsection{Radiation transfer}\label{sec:rad}

The typical luminosity of an AGB star is $\sim10^{37}$erg\ps. In our model, the AGB star's luminosity is $9.13\times10^{36}$erg\ps or $L=2342$L$_{\sun}$. This value is determined by calculating the radiation power at the hard pulsating sphere that underlies the photosphere i.e. $L=4\uppi\sigma_\text{sb}(0.9\text{au})^{2}\times(2900\text{K})^{4}$. If the secondary is a low-mass main-sequence star, its luminosity is $L_{\sun}$, which is small compared to the AGB star's luminosity. If the secondary is a white dwarf (WD), its luminosity will be close to the accretion luminosity. The upper limit of the accretion luminosity will be
\begin{equation}
    L_\text{acc}=214\frac{M_\text{WD}}{\text{M}_{\sun}}\frac{10^{9}\text{cm}}{r_\text{WD}}\frac{\dot{M}_\text{acc}}{10^{-7}(\text{M}_{\sun}\text{yr}^{-1})}\text{L}_{\sun}
\end{equation}
The maximum stable accretion rate measured in our calculation is $\sim1.2\times10^{-7}$\msun\pyr (table \ref{tab:masslosstable}) and the maximum secondary's mass is $0.5$\msun. Thus the accretion luminosity will be no greater than $200$L$_{\sun}$.  We note that any jets that form from the secondary will further decrease the available energy in luminosity \citep{soker1994}. With all these estimations, it is reasonable to consider the AGB star as the only radiation source in the binary system. Models that incorporates the feedback from the secondary can be studied in the future.

We apply 2D ray-tracing algorithm to calculate the radiation transfer. We notice that the binary system has periodic or symmetric outflows due to the circular orbital motion, so we average the radiation over the azimuthal angle where the axis of symmetry is crossing the centre of the AGB star and parallel to the z-axis. We assume single interacting photons in our calculations, i.e. photons will only interact with the dust or the gas once. In reality, there could be re-emitted photons and they could still interact with the dust or the gas but neglect that complication at present.

In equation (\ref{eqn:radforce}), the optical depth $\tau$ is calculated by:
\begin{equation}\label{eqn:tau}
    \tau(r_{0},\theta)=\int_{r_\text{ps}}^{r_{0}}\kappa(r,\theta)\rho(r,\theta)\text{d}r
\end{equation}
\begin{equation}
    \rho\left(r,\theta\right)=\frac{\int_{0}^{2\uppi}\rho\left(r,\theta,\phi\right)\text{d}\phi}{2\uppi}
\end{equation}
where $r_\text{ps}$ is the photosphere's radius, $r_{0}=\left|\bm{r}-\bm{r_\text{AGB}}\right|$, $\theta\in\left[0,\uppi\right]$ is the polar angle, $\phi\in\left[0,2\uppi\right]$ is the azimuthal angle and $\kappa=8$\cmspg if dust is present or $\kappa=2\times10^{-4}$\cmspg if dust is not present (see section \ref{sec:dust}).

\subsubsection{Dust formation and opacity}\label{sec:dust}

Dust can absorb and scatter photons effectively. When the photons leave the photosphere, they enter the optically thin regime and we use the Planck mean opacity calculated as \citep{chandrasekhar1950}:
\begin{multline}
    \kappa_\text{dust}\left(T_\text{bb}\right)=\\
    \frac{\int_{0}^{\infty}f\left(\lambda,T_\text{bb}\right)\left(\sigma_\text{abs}\left(\lambda\right)+\int_{0}^{\uppi}\sigma_\text{sca}(\lambda,\theta_\text{s})(1-\cos\theta_\text{s})\text{d}\theta_\text{s}\right)\text{d}\lambda}{m_\text{p}\int_{0}^{\infty}f\left(\lambda,T_\text{bb}\right)\text{d}\lambda}
\end{multline}
where $\kappa_\text{dust}\left(T_\text{bb}\right),f(\lambda,T_\text{bb}),\lambda,T_\text{bb},\theta_\text{s},\sigma_\text{abs},\sigma_\text{sca},m_\text{p}$ are the temperature-dependent Planck mean opacity of the dust, energy distribution of blackbody radiation, wavelength of the photon, blackbody temperature, scattering angle, absorption cross-section scattering cross-section and the mass of the scattering particle, respectively. The scattering for dust is forward scattering, that is, $\sigma_\text{sca}(\lambda,\theta_\text{s})$ approaches $0$ rapidly with the increase of $\theta_\text{s}$. Thus the scattering only contribute a small fraction of Planck mean opacity in dust and we can ignore it for simplicity.

The cross-section $\sigma_\text{abs}$ can be solved with Mie theory \citep{hulst1957,matzler2002} but it is dust species and dust size dependent. To carry out a quantitative study, we assume the dust is spherical and the size has the distribution of $\text{d}n\sim a^{-3.5}\text{d}a$ between a dust radius of $0.1-0.3\umu$m where $a$ is the dust radius and $n$ is the number density (\citet{mrn1977}, MRN distribution hereafter). We take the most common species of dust around oxygen-rich AGB stars (olivine and pyroxene \citet{jaeger1994,dorschner1995}).  In table \ref{tab:dustproperties} we present the Planck mean opacity for different temperature black bodys. 

\begin{table}
    \centering
    \begin{tabular}{c|c|c}
    \hline
    $T_\text{bb}$(K) &  $\kappa_\text{olivine}$(\cmspg)   & $\kappa_\text{pyroxene}$(\cmspg)\\ \hline
    2800 & 1787  & 461\\ \hline
    2600 & 1540 & 356\\ \hline
    2400 & 1303 & 269\\ \hline
    2200 & 1076 & 197\\ \hline
    2000 & 858 & 140\\ \hline
    \end{tabular}
    \caption{Planck mean opacity of olivine and pyroxene at different temperature of blackbody radiation. The energy distribution of the radiation is assumed to be in Planck distribution of that temperature. The dust grains are spherical and have MRN size distribution.}
    \label{tab:dustproperties}
\end{table}

The opacity is decreasing with the decreasing radiation temperature. Note that the black body radiation temperature is not necessarily equal to the equilibrium temperature of the dust. The latter is determined by the total flux of radiation while the former is not. In our simulation, the temperature at the photosphere is $2684$K ($T=\left(\frac{L}{4\uppi\sigma_\text{sb}(1.05\text{au})^{2}}\right)^{1/4}$), if we take the atmospheric reddening into consideration, the 'effective blackbody temperature' could be lower but that is beyond this paper's discussion. The total opacity of the gas-dust mixture is calculated by:

\begin{equation}
    \kappa_\text{total}=\frac{\rho_\text{gas}\kappa_\text{gas}+\rho_\text{dust}\kappa_\text{dust}}{\rho},
\end{equation}
where $\rho=\rho_\text{gas}+\rho_\text{dust}$ is the density used in the radiation transfer equations. $\kappa_\text{gas}=2\times10^{-4}$\cmspg, and $\rho_\text{dust}/\rho=0.01$ is used in our calculation \citep{bowen1988}. Given our assumptions, we choose $\kappa_{total}=8$\cmspg. This value may be slightly larger than the real opacity of gas-dust mixture around an AGB star but it lies within a reasonable range for the calculations.

The dynamic dust formation is also included in our simulation. Dust forms when the equilibrium temperature is below the dust's condensation temperature; we calculate the equilibrium temperature by

\begin{equation}
    T_\text{eq}=\sqrt[4]{\frac{\beta L \exp\left(-\tau\right)}{4\uppi\sigma_\text{sb}r^2}}
\end{equation}

Where $\beta=0.3,\sigma_\text{sb},r$ are the coefficient of absorption, Stefan-Boltzmann constant and the distance to the centre of the AGB star respectively. $\tau$ will be calculated by equation \ref{eqn:tau}. We add $\beta$ in our equation because the dust can only absorb short and specific wavelength of radiation. We choose this value because it gives us reasonable AGB winds (section \ref{sec:isoagb}). Finally, the condensation temperature of dust is assumed to be $1500$K.

\subsubsection{Cooling}\label{sec:cooling}

A very important physical ingredient in binary simulation is cooling. The accretion on to the secondary star and the strong shocks around the AGB star will compress and heat the gas. Disc formation and accretion rates are dependent on cooling \citep{lee2015}.

We use the same cooling mechanisms as \citet{mastrodemos1998}: electron excitation of \ion{H}{i} and molecular cooling between molecular hydrogen and water. We updated a few input parameters to make the cooling more realistic.

When outside the photosphere and when $T>4000$K, the electron excitation of \ion{H}{i} is considered. The cooling rate is calculated by $\dot{Q}=7.3\times10^{19}n_\text{e}n_\text{H}\exp\left(-118400/T\right)$ \citep{spitzer1978}. The number density of electrons is characterized by a step function to further distinguish high temperature cooling and low temperature cooling. For $4000<T<8000\text{K}$, $n_\text{e}=6.0\times10^{-4}n_\text{H}$ and for $T>8000$K, $n_\text{e}=1.0\times10^{-3}n_\text{H}$.

Molecular cooling is more complicated. In reality, the formation of molecules requires both low temperature and consideration of relaxation times. H$_2$ is scarce in the constantly shocked AGB atmosphere since the timescale of H$_2$ formation is longer than the time between successive shocks \citep{bowen1988,cherchneff2006}. We empirically choose the H$_2$ formation region to be close to the dust formation region, thus molecular cooling will not be considered within a $1.8$au radius of the AGB star. Outside the $1.8$au sphere, when $T<4000$K, the molecular cooling is calculated by following \citet{nk1993} and \citet{mastrodemos1998}.

\subsection{Isolated AGB star}\label{sec:isoagb}

\begin{figure}
    \centering
    \includegraphics[width=1.0\columnwidth]{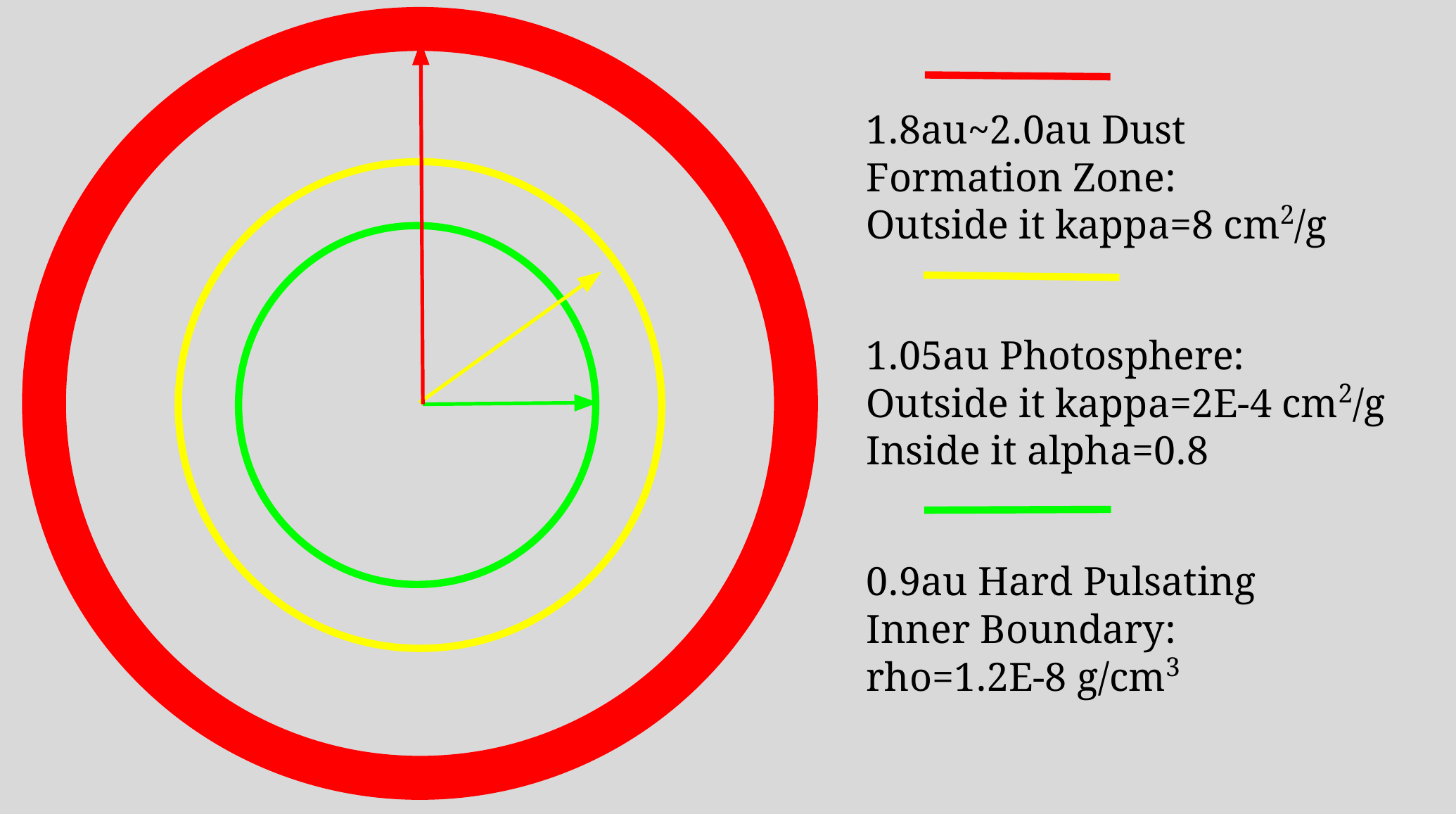}
    \caption{The AGB star's structure in our model.}
    \label{fig:agbstructure}
\end{figure}

Before discussing the binary model we first describe our representation of the isolated AGB star. The test results of normal pulsating AGB star and the AGB star with burst will be presented in separate subsections. Our isolated AGB star model is phenomenological and mainly inspired by \citet{bowen1988} and \citet{liljegren2016}. A sketch of the AGB star's structure is shown in Fig. \ref{fig:agbstructure} and its physical properties are listed in Table \ref{tab:isolatedstar}.

There are three important shells in the isolated AGB star model: the 'hard' pulsating inner boundary, the photosphere and the dust formation zone. The radius of hard pulsating inner boundary is the hydrodynamic boundary fixed in radius, density and temperature but varying in velocity. It mimics the convective instability in the atmosphere \citep{freytag2008} and drives shocks (or potentially a burst) outwards. This boundary is 'hard' because it will not be affected by the condition on the outside and therefore the dynamics of this shell are predetermined. The photosphere is presumed to be shell with inner edge $r=1.05$au. In reality, the photosphere boundaries should be dynamic but that requires a more detailed radiation transfer calculation (and computational resources) than we attempt here. Between the pulsating boundary and the photosphere we use a reduced gravity described as
\begin{equation}
    \bm{f}_\text{gAGB}=\alpha\frac{G M_\text{AGB}}{r_\text{AGB}^{2}}\hat{\bm{r}}_\text{AGB},
\end{equation}
where $\alpha=0.8$. This reduced gravity could come from a combination of radiation pressure and instabilities. In our model testing experience, different $\alpha$ value will give us different wind properties and there is no simple linear relation of $\alpha$ to wind speed and mass-loss rate. For example, in the small $\alpha$ situation, the pulsation at the hard pulsating sphere may push more material to the photosphere and dust formation zone with higher speed but the optical depth will increase dramatically at the dust formation zone. Then the radiation pressure may drop fast with the increase of radii and the wind may decelerate. How useful is $\alpha$ and how to justify $\alpha$ more rigorously remains a question in phenomenological AGB wind models. We also allow for the atmospheric reddening in our binary model using $\kappa_{gas}=2\times10^{-4}$\cmspg between the pulsating boundary and the outer edge of photosphere. Outside the photosphere, the gas opacity is kept the same with $\alpha=1.0$ and the cooling from electron excitation turned on. Finally, we have the dust formation zone whose inner edge is dynamic due to the atmospheric reddening. In the 3D isolated AGB star simulation, the dust formation radius is between $1.8\text{au}-2.0\text{au}$.

\begin{table}
    \centering
    \begin{tabular}{||c|c||}
    \hline
        Mass & $1\text{M}_{\sun}$ \\ \hline
        Luminosity & $2342\text{L}_{\sun}$\\ \hline
        Pulsating boundary density & $1.2\times10^{-8}$gcm$^{-3}$\\ \hline
        Outflow temperature & $2900$K \\ \hline
    \end{tabular}
    \caption{AGB star's physical properties.}
    \label{tab:isolatedstar}
\end{table}

We now  present the two AGB sub-models used in the simulations.

\subsubsection{Pulsation}\label{section:pulsation}

\begin{figure}
    \centering
    \includegraphics[width=1.0\columnwidth]{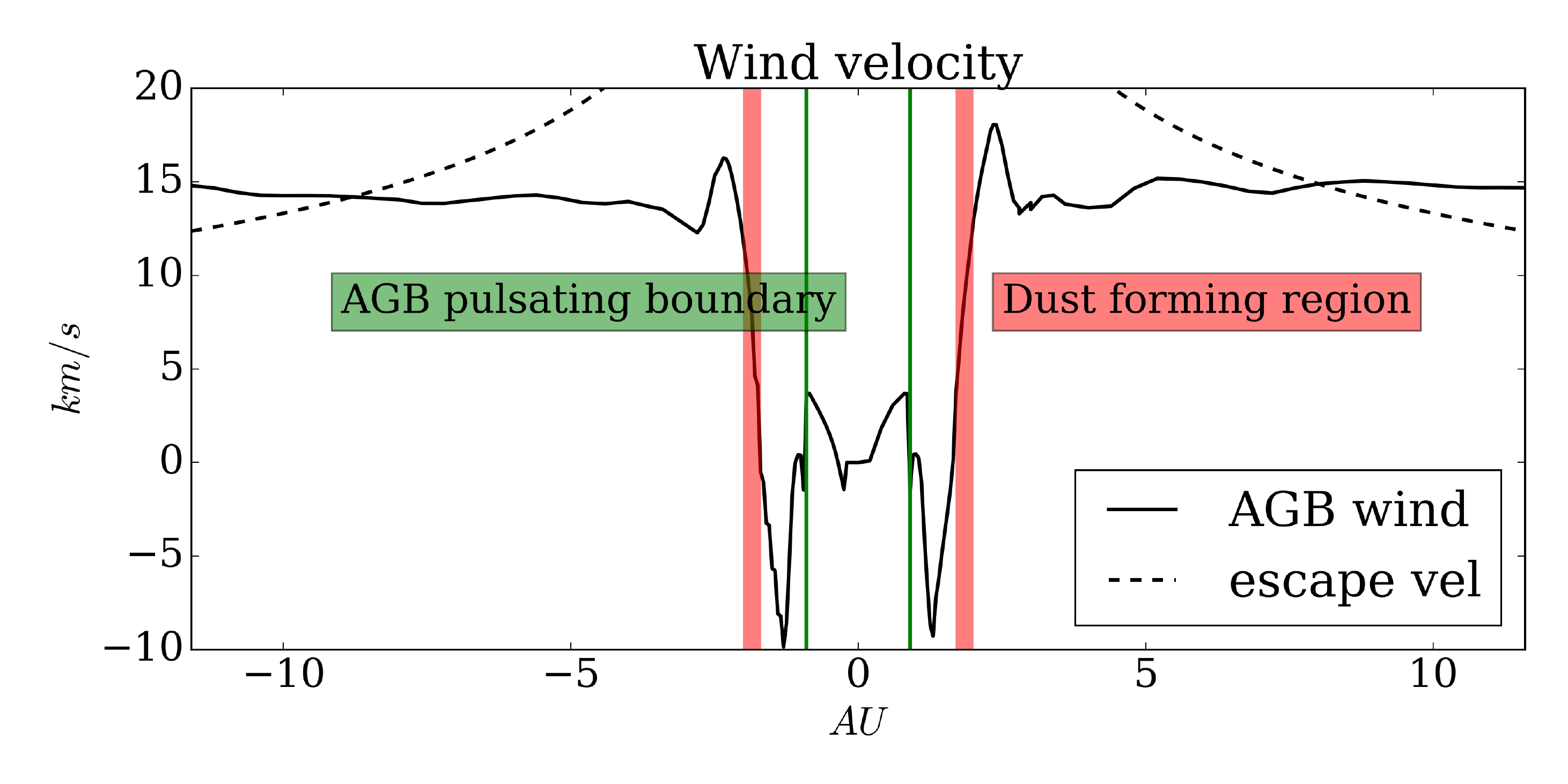}\\
    \includegraphics[width=1.0\columnwidth]{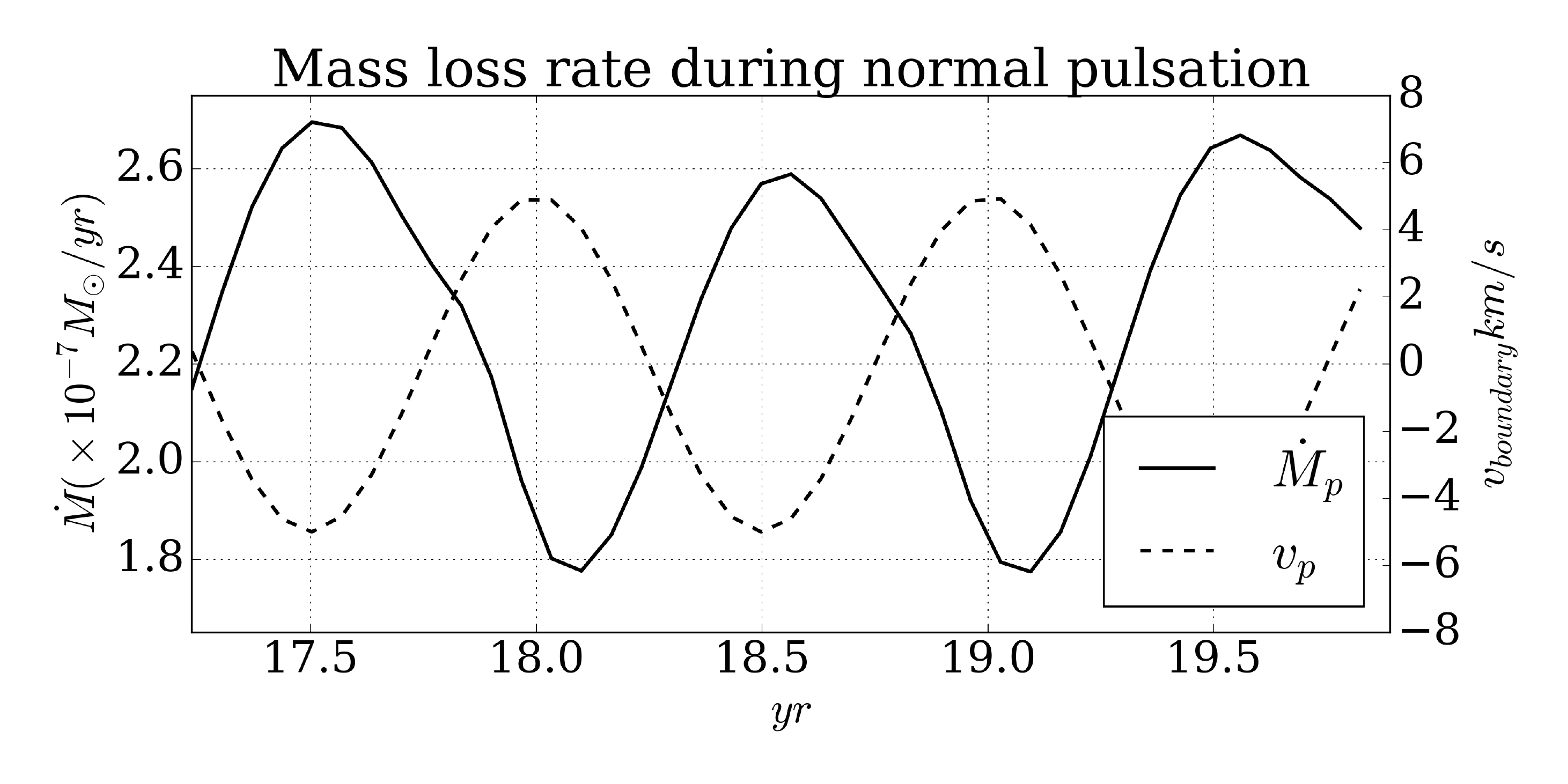}
    \caption{The upper panel shows the velocity profile of the AGB wind. The centre of the AGB star is located at the origin. The red strips illustrate the dust forming region and the green strips illustrate the boundary of the AGB star. Positive value means the velocity is in outward direction of the star and negative value means the velocity is in inward direction of the star. The lower panel shows the mass loss rate (solid line and left $y$-axis) and the boundary velocity (dashed line and right $y$-axis) of the AGB star during the normal pulsating periods as a function of time.}
    \label{fig:isolatedpulse}
\end{figure}

We will always assume pulsation in our binary simulations and we use it to solve the isolated spherically symmetric pulsating AGB wind problem in 3D with {\small ASTROBEAR}. The physical dimension of the simulation box is $(24\text{au}^3$. The base resolution is $240^3$ and we add one level of AMR; therefore, the physical dimension of smallest cell is $(0.05\text{au})^3$. The numerical solution for the outflow from one of our pulsating isolated AGB stars is shown in Fig. \ref{fig:isolatedpulse}.

The upper panel of Fig. \ref{fig:isolatedpulse} shows that  the wind velocity approaches $15$km\ps asymptotically. The flatness of the velocity profile implies that the radiation pressure on the wind  balances the gravitational force. This balancing is dynamic because the optical depth (Eq. \ref{eqn:tau}) changes within one period of pulsation. Thus the radiation force (Eq. \ref{eqn:radforce}) also changes. Between the surface of the pulsating boundary and the dust forming region, the velocity can go negative, so fall back can occur. In this region, gas is constantly shocked as material rises and falls. 
The escape velocity to the AGB star's gravity is shown with dashed line. 

In the lower panel of Fig. \ref{fig:isolatedpulse} we see the radial velocity at the pulsating boundary which is $v_{p}=5\cos{\left(2\uppi t/P\right)}$\text{km}\ps where $P=365$d. The mass loss rate is measured at $10$au, that is, we summed the mass flux through the $10$au shell. Thus we expect to see the mass loss rate out of phase with the pulsation. We can however see from Fig. \ref{fig:isolatedpulse} that the mass loss rate has the same period as the pulsation. The average mass loss rate during normal pulsation is $\dot{M}_\text{avg}=2.31\times10^{-7}$\msun\pyr. This is smaller than the mass loss rate predicted by the empirical formula given by \citet{vanloon2005}. Their formula predicts a mass loss rate of $2.57\times10^{-6}$\msun\pyr. Our lower mass loss rate is mainly a consequence of not constructing a fully self-consistent atmosphere which would at least require a more accurate equation of state and radiation transfer. Enough resolution should also be contained within each scale height. Since the intention of our present paper is to study wind binary interactions we only need to get the general characteristics of the wind correct (dust driving etc).

\subsubsection{Burst}\label{sec:busrt}

In our isolated AGB star model, a 'burst' is an increase in mass outflow over a short period of time of the AGB star. In reality, the outflow could be induced by the in plunge of a planet with $M_\text{p}\in\left[M_\text{J},50M_\text{J}\right]$ \citep{nordhaus2006,nordhaus2007,nordhaus2013}. 3D hydrodynamic calculation \citep{staff2016b} shows that a $10M_{J}$ planet could unbind $10^{-4}-10^{-3}$\msun of the giant star's envelope and increase the giant's luminosity. \citet{ivanova2016} use 3D hydrodynamic model and shows that a $0.05$\msun planet could unbind $0.02$\msun of the giant's envelop. They both see 'fallback' phenomenon \citep{chen2016b} in the simulations. In the hope of testing the dynamics of such burst in binary systems, we build our phenomenological model to mimic their solutions. The details of plunge in process of a planet is not included in our phenomenological model, thus the 'burst' exists numerically. Next we will explain how we induce the 'burst' in our isolated AGB star model. It remains the same in the binary simulation.

\begin{figure}
    \centering
    \includegraphics[width=1.0\columnwidth]{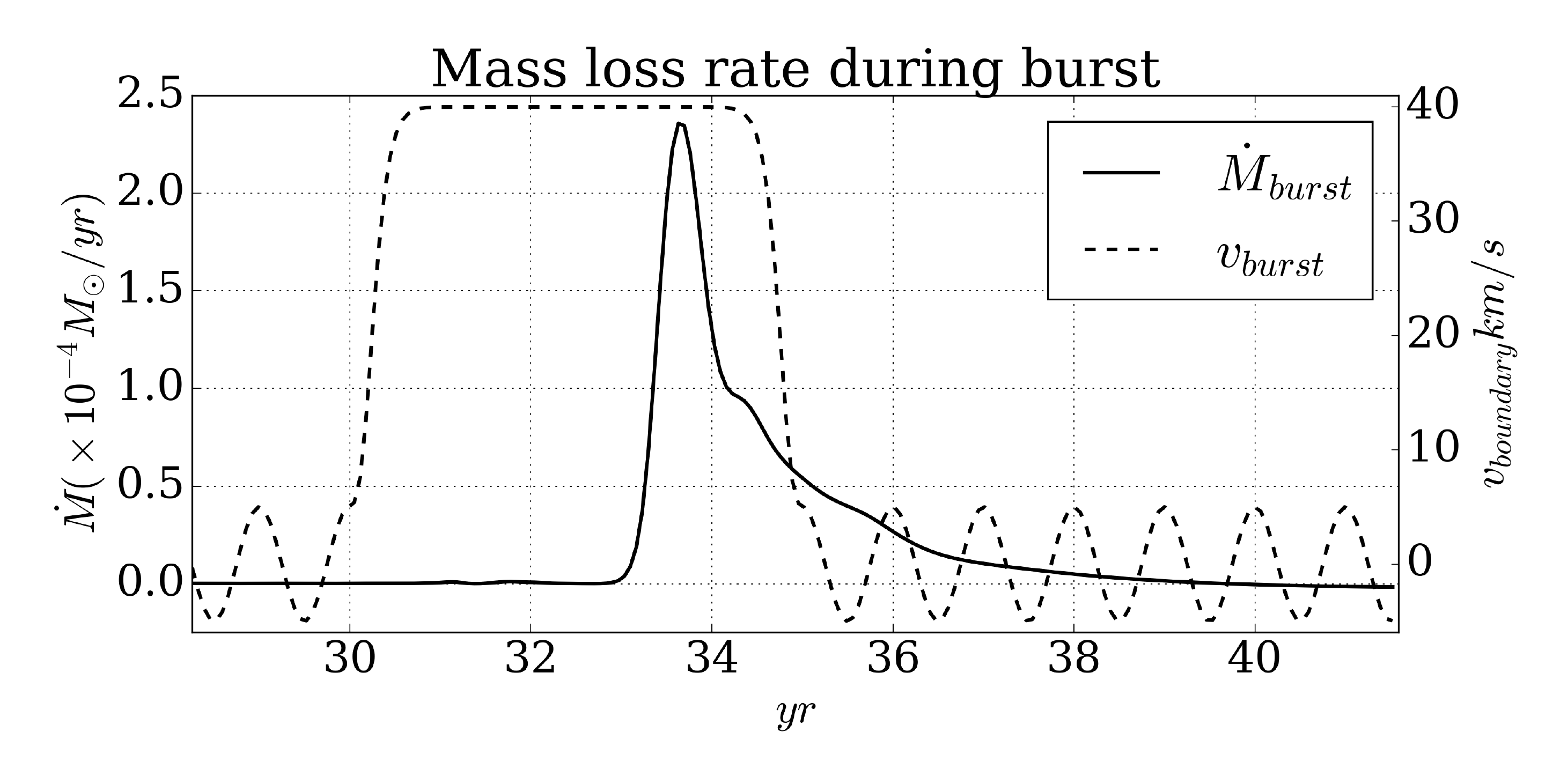}
    \caption{Mass loss rate and the boundary velocity of the AGB star during the burst event as a function of time.}
    \label{fig:burst}
\end{figure}

The total span of the burst is $5$yr. During the burst, the outflow density at the hard pulsating sphere will change from $1.2\times10^{-8}$ to $1.2\times10^{-12}$\gcc, the outflow velocity which also lies at the hard pulsating sphere will vary with time:
\begin{equation}
    \frac{v_\text{burst}}{\text{km }\text{s}^{-1}}=5+35\exp\left(-\frac{((t/\text{yr}-32.5)/2)^{20}}{(5/2.2)^{20}}\right),
\end{equation}
where $30<t<35\text{yr}$. This expression is chosen to be generic and any smooth function with similar shape would work. The outflow temperature will increase from $2900$ to $3450$K and the luminosity will be doubled to $4684$L$_{\sun}$. We also stretch the reduced gravity region to $2$au from $1.05$au. Our goal in driving the burst is to have an outflow that captures some of the key features of what would be found in more from more detailed models. The hydrodynamics deep inside the burst is not our primary interest and we chose parameters that heuristically allow the wind to increase its mass lose rate. Fig. \ref{fig:burst} presents the numerical solution of the burst.

Like the pulsating AGB model, the burst model is also solved with {\small ASTROBEAR} in 3D. The burst lasts from $30$ to $35$yr. After the burst, the AGB star returns to the normal pulsating state. We still measure the mass loss rate at $10$AU, so  there is lag between the burst and the peak of the mass loss. The mass-loss rate ramps up to the peak ($2.3\times10^{-4}$\msun\pyr) rapidly then decays and finally some leftover material starts to fall back. The total mass loss driven by the burst is roughly $2.9\times10^{-4}$\msun.

\subsection{Accreting star}

The accreting secondary is the other important object in binary simulations. We first consider the range of physics which might be included in including the secondary.  As mentioned before, a low-mass-main-sequence stellar accretion disk is not luminous, but it might have an extended magnetic field. In that case, the magnetic field and the disk could interact with each other. If the secondary is a WD, the radiation from the disk could be strong enough to change the dynamics of the accretion process, especially for  possibly super-Eddington accretion in the RLOF. The temperature of the innermost region of the disk can be up to $\sim10^{6}$K and the lowest temperature in an AGB binary system can be $\sim10$K. Thus one needs a continuous tablular equation of state from $10-10^{6}$K \citep{pejcha2016,tomida2015}. The density is also spans many orders of magnitude ($10^{-20}-10^{0}$\gcc). The possible emerging outflows could change the morphology of the binary system and the accretion disk \citep{blackman2001,frank2004,blackman2009}.  Diskoseismology \citep{nowak1991} is relevant if self-gravity is to be considered in the binary simulation. All these questions converge to the question of how to treat the inner boundary condition of the accreting star.

Given resolution constrains (even with AMR) in our calculation the accreting star is a gravitating sink particle and we adopt the \citet{krumholz2004} prescription. The sink particle has been tested via the BH accretion problem where the accretion rate, numerical viscosity and the accretion flow under various conditions were examined and found to produce excellent agreement with analytic results. One must be careful in that this prescription for sink particles can produce errors in the accretion rate when the flow is transitioning from subsonic to supersonic at the accretion radius.

\subsection{Outer boundary condition}

For our outer boundary we carve out a cylinder with its axis of symmetry parallel to the z-axis inside the cuboid.  The radius of the cylinder is one cell smaller than the half length of the x or y dimension. In the lab-frame, we let the region  outside the cylinder have high Mach number supersonic  outward flows. The outward flow condition is reset before every step of the simulation. Then we transform the Galilean variant quantities ($\bm{u},E$) of the outside cylinder region to the corotating frame. The advantage of this boundary condition is that it prevents unphysical inflow from entering the grid due to the frame rotation. For the top and bottom surfaces of the cylinder, the physical quantities in the corotating frame ($\rho,\bm{u},p,E,e,T$) outside the boundaries are first extrapolated. The perpendicular component of the velocity is then calculated as
\begin{equation}
    u_\text{p}=\bm{u}\cdot\hat{\bm{n}}_\text{b},
\end{equation}
where $\hat{\bm{n}}_{b}$ is the unit outward normal vector of that boundary. If $u_{p}\ge0$, we  do nothing; if $u_{p}<0$, we set it to $0$. This prescription  eliminates  unphysical fallback of  material into the simulation.

\section{binary simulations}\label{sec:binary}

\subsection{Numerical setup of binary simulations}

The setups of the six runs are listed in table \ref{tab:modellist}. The column named \textbf{AMR} represents the highest level of AMR in the simulation. This highest  level  is  only applied to the region around the secondary in the 'no burst' models. The AGB star is kept at  the same resolution as in the isolated model. We do this to focus the computational power on  resolving the accretion disk around the secondary to the finest level. For the binary model with the burst, we do not use the same level of AMR as model 2 so that we can compare the accretion rate before burst to model 2. We find that their accretion rate is nearly the same, this feature ensures that our resolution around the secondary does not influence the accretion rate, i.e. we have enough resolution.

\begin{table*}
    \centering
    \begin{tabular}{||c|c|c|c|c|c|c|c|c|c|c|c||}
    \hline
    model & $M_\text{AGB}$ & $M_\text{sec}$ & $d$ & $d_{x},d_{y}$ & $d_{z}$ & $N_{x},N_{y}$ & $N_z$ & AMR & Burst & $r_\text{acc}$ & $r_\text{flux}$ \\
    & \msun & \msun & au & au & au & & & & & au & au\\\hline
    1 & 1.0 & 0.1 & 3 & 40 & 40 & 100 & 100 & 4 & No & 0.1 & 3 \\ \hline
    2 & 1.0 & 0.5 & 4 & 96 & 48 & 60 & 30 & 6 & No & 0.1 & 4 \\ \hline
    3 & 1.0 & 0.5 & 6 & 80 & 40 & 100 & 50 & 5 & No & 0.1 & 7 \\ \hline
    4 & 1.0 & 0.5 & 8 & 128 & 48 & 80 & 30 & 6 & No & 0.1 & 7 \\ \hline
    5 & 1.0 & 0.5 & 10 & 192 & 64 & 60 & 20 & 7 & No & 0.1 & 9 \\ \hline
    6 & 1.0 & 0.5 & 4 & 96 & 48 & 60 & 30 & 5 & Yes & 0.2 & N.A. \\ \hline
    \end{tabular}
    \caption{The first column lists the model number used throughout this paper; $d(\text{au})$ is the binary separation; $d_{x},d_{y}(\text{au})$ and $d_{z}(\text{au})$ are the physical dimensions of x,y and z of the simulation box and we take $d_{x},d_{y}(\text{au})$ for all simulations; $N_{x},N_{y}$ and $N_z$ are the number of base cells in x,y and z dimensions. The \textbf{AMR} column lists the highest level of AMR for each model. \textbf{Burst} refers to the bursting state of the AGB star in section \ref{sec:busrt}. Only the AGB star in model 6 has a burst. $r_\text{acc}$ is the accretion radius of the sink particle. $r_\text{flux}$ is the radius of the shell through which the escaping flux is measured.}
    \label{tab:modellist}
\end{table*}

We now explain the models in the table by taking model 1 as an example. For model 1, we simulate a binary system with $1M_{\sun}$ AGB star and a $0.1\text{M}_{\sun}$ companion. The separation of the binary stars is $3$au. The physical dimension of the simulation box is $(40\text{au})^3$. We divide the space into $100^3$ base cells so that each base cell has a physical volume of $(0.4\text{au})^3$. With 4 levels of AMR, the physical volume of the finest cell is $(0.025\text{au})^3$. The region with highest level of AMR is restricted to the Roche lobe of the secondary. The resolution of the AGB star is the same as in section \ref{section:pulsation}, i.e. the physical volume of the finest cell is $(0.05\text{au})^3$. It is assumed that the secondary will accrete the gas within $r_{acc}=0.1$au. The AGB star is  in its normal pulsating state throughout the simulation.

We do not deduct the mass lost from the AGB wind from the mass of the star in our calculation because the total mass loss in the entire simulation is small compared to the mass of the AGB star. The total mass loss $\text{d}M$ in a $100$yr simulation is $< 3\times10^{-5}\text{M}_{\sun}$. This value will not exceed $3\times10^{-4}\text{M}_{\sun}$ even in the simulation with burst. We expect that such a small mass change will not signficantly alter the dynamics of the AGB wind, the accretion of the secondary, or the outflow morphology or  orbital dynamics. For the same reason, the angular frequency in the corotating frame simulations and the center of mass are fixed.

For  longer simulations (e.g. $>>100$yr) than considered here however,  the mass loss will be important over the lifetime of the AGB star ($M_\text{AGB}/\dot{M}_\text{AGB}$) and is a topic of  future work.

\subsection{Dynamics of the mass loss rate and the accretion rate}\label{sec:affmlrateandaccrate}

In our calculations, we find  that the secondary influences the mass loss rate from the primary and that the accretion rate on to the secondary varies dramatically with different binary separations, $d$. The results are listed in table \ref{tab:masslosstable} where $\zeta=\frac{\dot{M}_\text{acc}}{\dot{M}_\text{AGB}}$ and $\epsilon=\frac{\dot{M}_\text{AGB}-\dot{M}_\text{avg}}{\dot{M}_\text{avg}}$.

\begin{table}
    \centering
    \begin{tabular}{||c|c|c|c|c||}
    \hline
        model\# & $\dot{M}_\text{AGB}$ & $\dot{M}_\text{acc}$ & $\zeta$ & $\epsilon$\\
        & \msun\pyr & \msun\pyr & per cent & per cent\\\hline
    1 & $2.960\times10^{-7}$ & $1.159\times10^{-7}$ & $39.2$ & $28.1$ \\ \hline
    2 & $3.325\times10^{-7}$ & $1.224\times10^{-7}$ & $36.8$ & $43.9$ \\ \hline
    3 & $2.208\times10^{-7}$ & $4.513\times10^{-8}$ & $20.4$ & $-4.42$ \\ \hline
    4 & $2.658\times10^{-7}$ & $8.097\times10^{-9}$ & $3.05$ & $15.1$ \\ \hline
    5 & $2.606\times10^{-7}$ & $5.848\times10^{-9}$ & $2.24$ & $12.8$ \\
    \hline
    \end{tabular}
    \caption{Model numbers and the  average mass loss rate of the AGB star and the mass accretion rate of the secondary. We denote the percentage of the accretion rate to the mass loss rate as $\zeta$. We also show the percentage $\epsilon$ of the changed mass loss rate to the mass loss rate in isolated AGB model.}
    \label{tab:masslosstable}
\end{table}

How do we calculate $\dot{M}_\text{AGB}$? The mass lost from the AGB star can go to  each of four places:
\begin{enumerate}
    \item It can escape the binary system at rate of $\dot{M}_\text{esc}$.
    \item  It can fall back into a  circumbinary disk $(\dot{M}_\text{cd})$
    \item It can fall into an accretion disk around the secondary ($\dot{M}_\text{ad}$).
    \item It can fall directly on to the secondary and be accreted at rate of $\dot{M}_\text{acc}$.
    \end{enumerate}

The mass loss rate of the AGB star is thus the sum of these four contributions:
\begin{equation}
    \dot{M}_\text{AGB}=\dot{M}_\text{esc}+\dot{M}_\text{cd}+\dot{M}_\text{ad}+\dot{M}_\text{acc}
\end{equation}
When the simulation has come to stable state,  the mass in the circumbinary disk and in accretion disk around the secondary are nearly constant so that  ($\dot{M}_\text{cd}\approx0$ and $\dot{M}_\text{ad}\approx0$). Therefore:
\begin{equation}
    \dot{M}_\text{AGB}\approx\dot{M}_\text{esc}+\dot{M}_\text{acc}
\end{equation}
The accretion rate $\dot{M}_\text{acc}$ is calculated by {\small ASTROBEAR} using the prescription introduced by \citet{krumholz2004} when the simulation is in stable state. The detail algorithm can be found in section 2.4 in \citet{krumholz2004}. When the simulation has reached its stable state,$\dot{M}_\text{acc}$ does not vary much (see the first 30 years in Fig. \ref{fig:accrete}) during pulsation but we still take the average value over 10 pulsation period. $\dot{M}_\text{esc}$ is calculated by summing the flux through a shell of radius $r_\text{flux}$ (see table \ref{tab:modellist}) centered at the origin. Since the flux and the accretion rate  vary with the pulsation, we take the average value over 10 pulsation periods when the binary system has come to stable state.

As defined above, $\epsilon$ measures the change in mass loss rate compared to the isolated AGB model. It is evident from the table that the AGB mass loss rate is different when there is a companion. There are at least three ways that the secondary can affect $\dot{M}_{AGB}$:
\begin{enumerate}
    \item The secondary can attract material around it to make the optical depth rise. The reduces the radiation force on the radiation-shielded gas and some of it will fall back toward the AGB star.
    \item The secondary may not be able to accrete material at its inner boundary  as fast as it accretes material at the outer boundary of its disk. Material will then pile up in the accretion disk. Some of it will flow through the L2 point \citep{pejcha2016} and leave the system.
    \item The secondary can pull some of the material inside the dust fomrmation radius (i.e. before it is radiatively accelerated)  that would otherwise fall back on to the AGB star in the isolated model.
\end{enumerate}
The first mechanism above (optical depth mechanism)  reduces $\dot{M}_\text{AGB}$ while the second (L2 mass loss mechanism) and the third mechanisms  increase $\dot{M}_\text{AGB}$. In model 1, 2 and 3, all three mechanisms operate; L2 and L3 mass loss are found in model 1 and 2 and L2 mass loss is found in model 3 (Fig. \ref{fig:binarysimulation}). For model 4 and 5 the third mechanism above dominates via BH accretion on to the secondary.

The accretion rate varies by a factor of $20$ between different models. In fact the accretion rate can be up to $\zeta=39.2$ per cent (model 1) of $\dot{M}_\text{AGB}$ in our simulations. In that model, the dust formation radius is $r_\text{dust}\approx1.8$au. The pulsation makes the optical depth oscillate and thus the  radiation force from the AGB star  also oscillates. The AGB star's Roche lobe  deforms  and can detach from that of the secondary under this oscillating radiation force and the L1 point may vanish in some circumstances \citep{dermine2009}. Given that the typical BH accretion gives $\zeta\ll10$ per cent whereas RLOF can produce $\zeta\approx100$ per cent, we interpret this high but not $100$ per cent mass transfer efficiency to exemplify WRLOF. For models 4 and 5 respectively, $\zeta=3.25$ per cent and $\zeta=2.24$ per cent. These values and the spiral outflow wind  from the binary (Fig. \ref{fig:binarysimulation}) and suggest that  models 4 and 5 exhibit BH accretion.

\subsection{Results from the 'no burst' binary simulations}

\begin{figure}
    \centering
    \includegraphics[width=1.0\columnwidth]{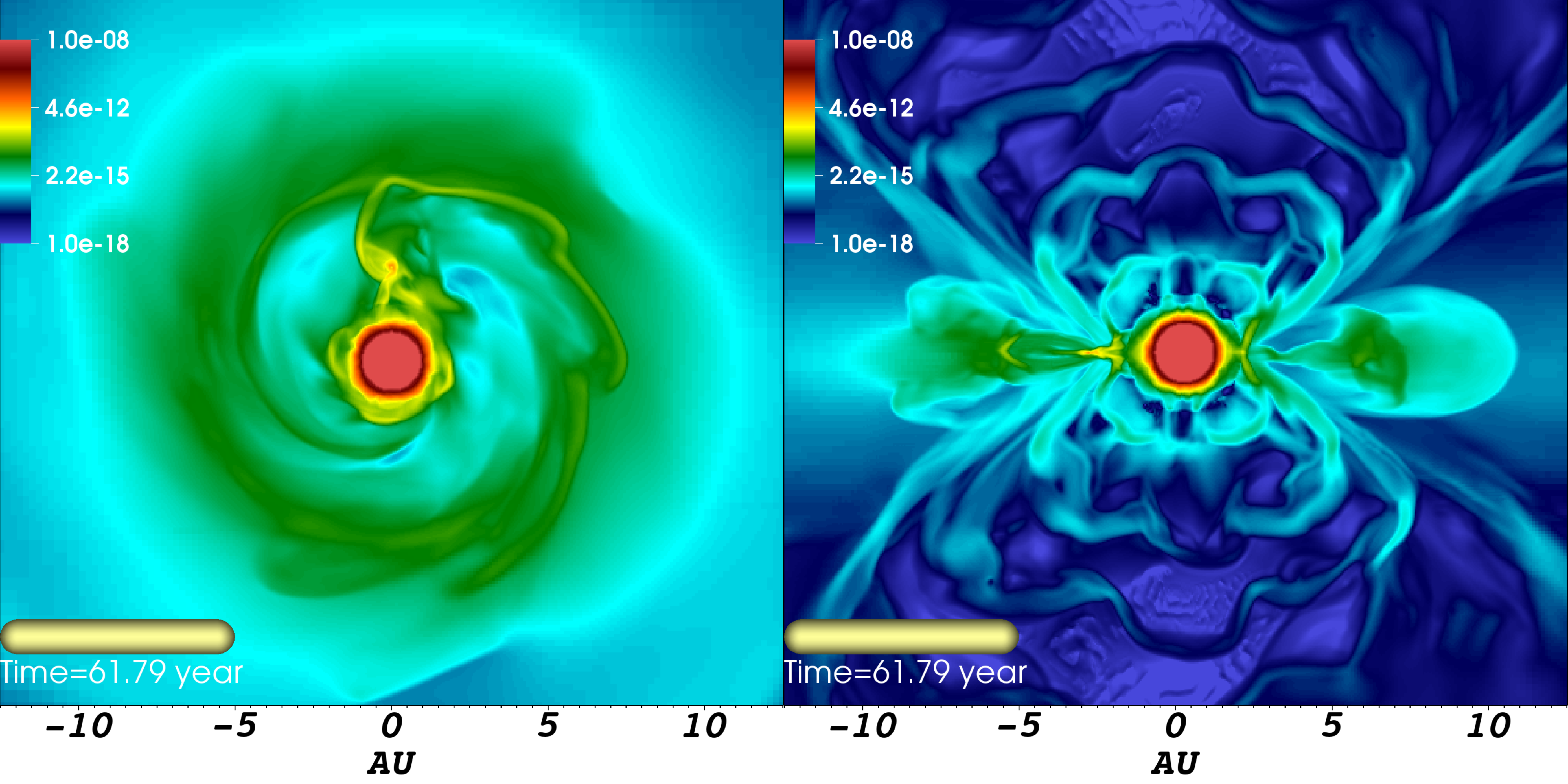}\\
    \includegraphics[width=1.0\columnwidth]{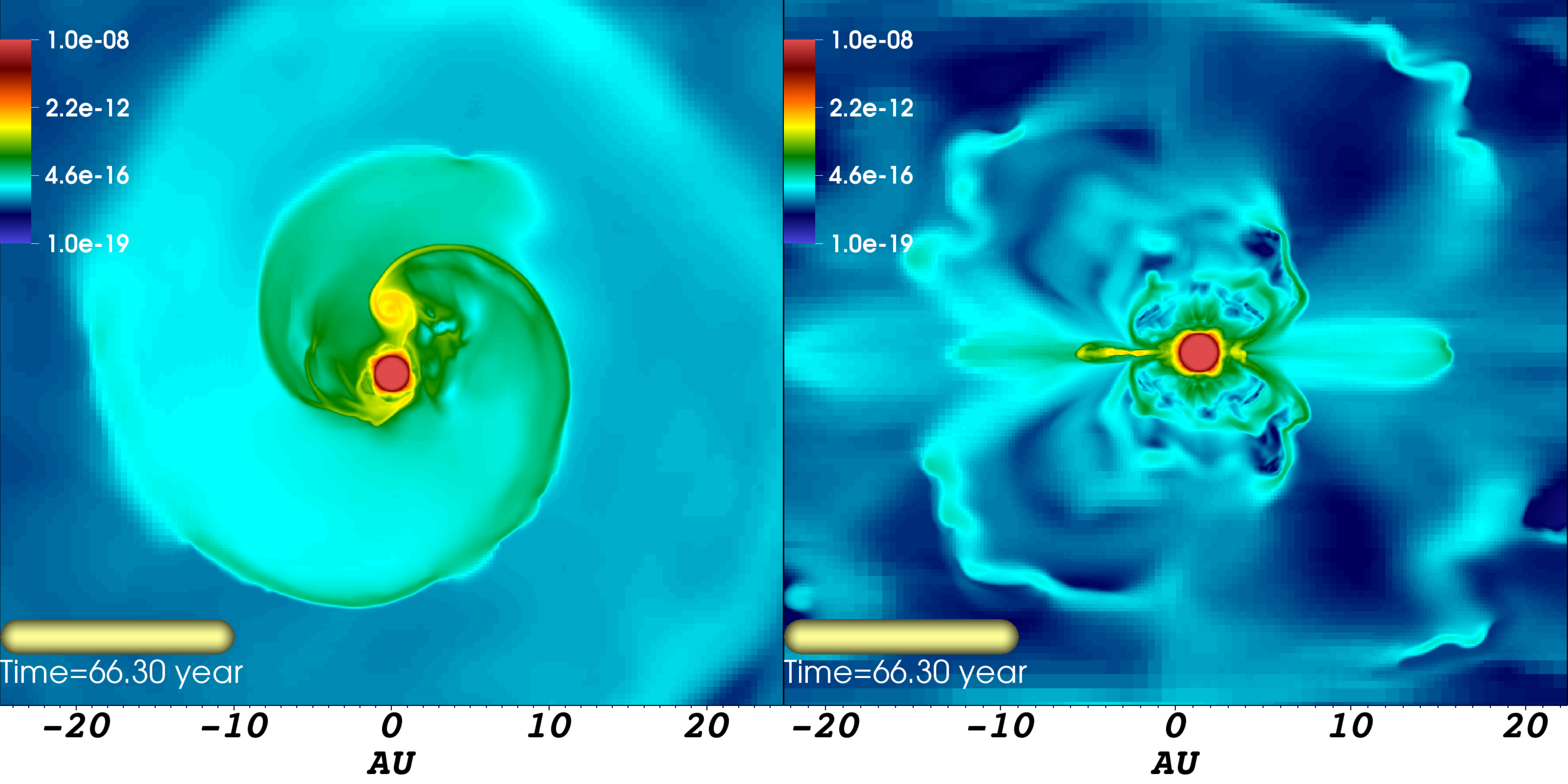}\\
    \includegraphics[width=1.0\columnwidth]{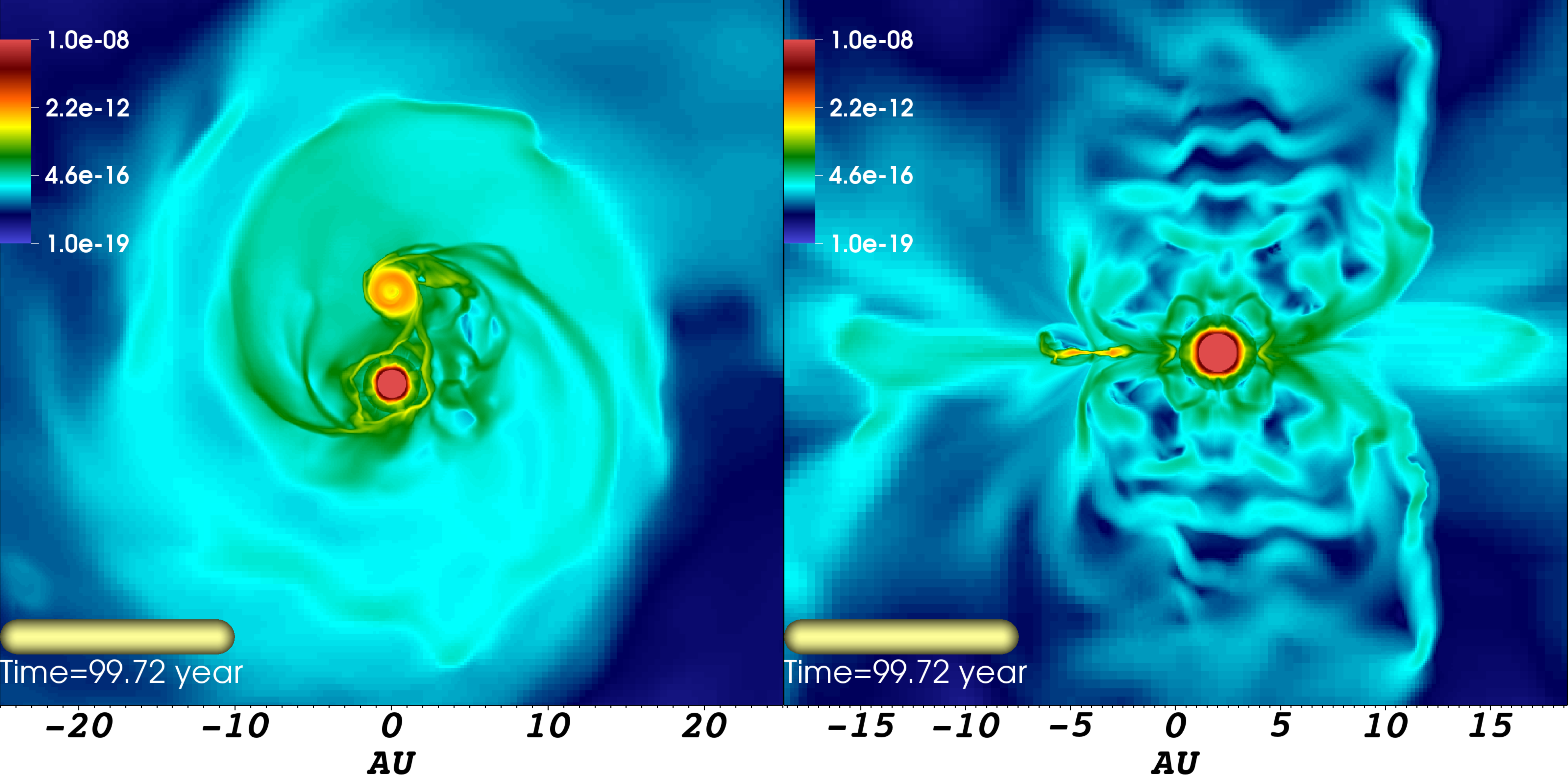}\\
    \includegraphics[width=1.0\columnwidth]{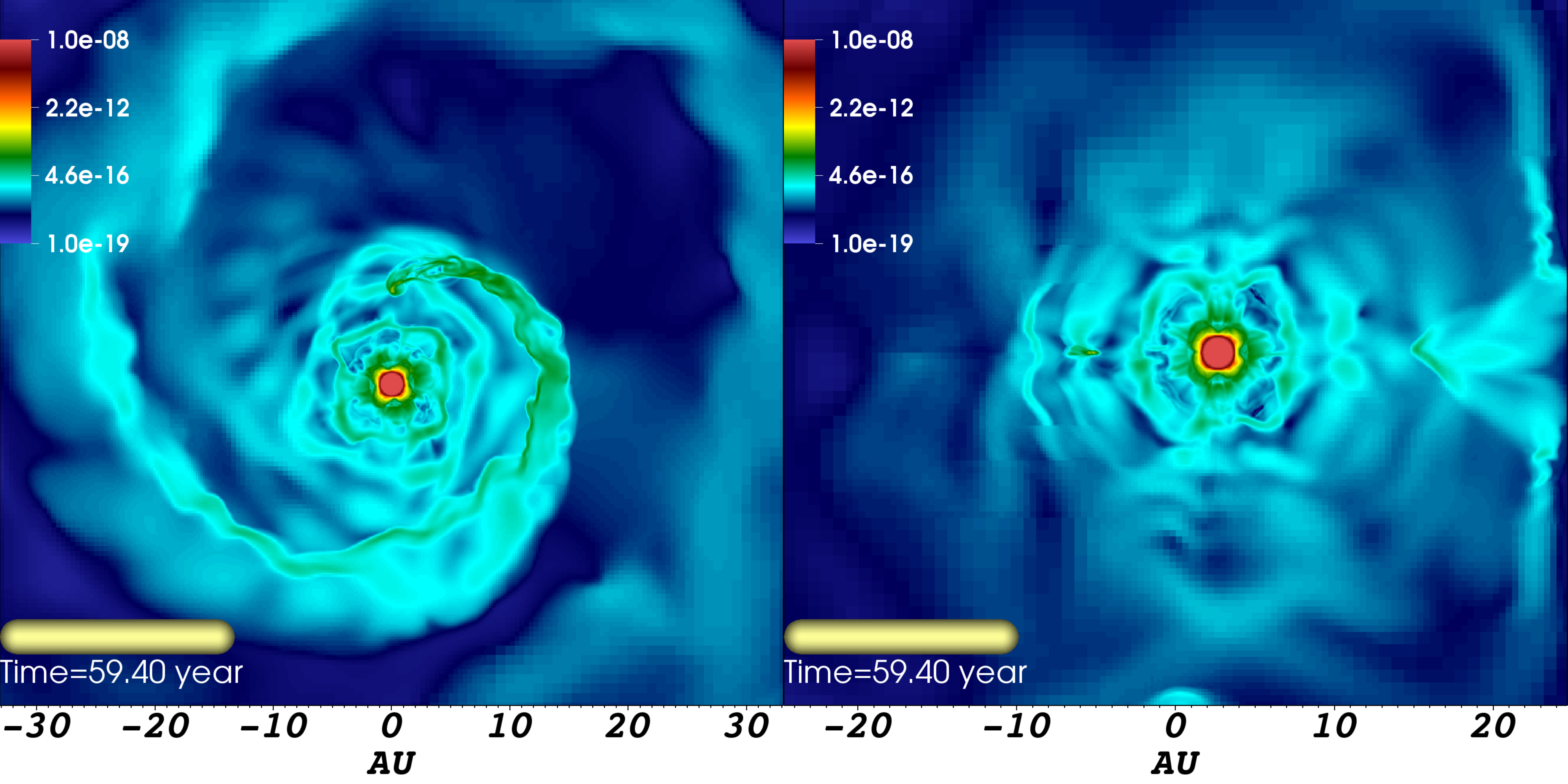}\\
    \includegraphics[width=1.0\columnwidth]{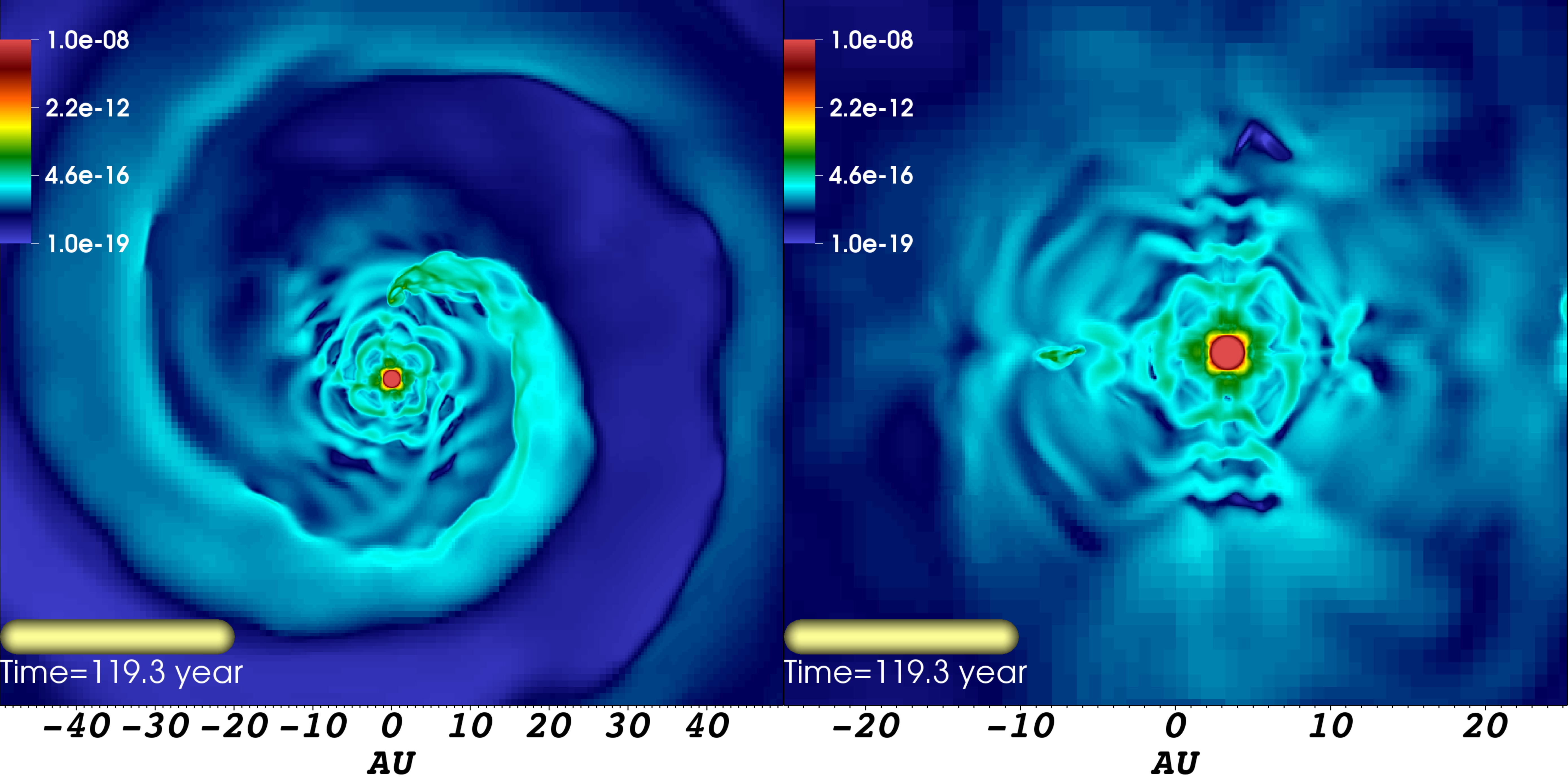}
    \caption{From top to bottom, figures in nth row correspond to the nth 'no burst' model listed in table \ref{tab:modellist}. Figures in the left column show the density  in \gcc at the $z=0$ plane cut (also the equatorial plane). Figures in the right column show the $x=0$ mid-plane  cut.}
    \label{fig:binarysimulation}
\end{figure}

We present the figures from model 1 - model 5 in Fig. \ref{fig:binarysimulation}. The figures show results from  simulations that which achieve a stable state for at least $10$yr.

\subsubsection{Bipolar outflow}

The low mass secondary in orbit around  AGB star can shape the AGB wind into a bipolar outflow structure \citep{nordhaus2006}. An observational example is L2 Puppis \citep{kervella2015,kervella2016}. \citet{chen2016a} found that  3D hydrodynamic simulation of  L2 Puppis and found very good agreement with  the morphology and  spectral energy distribution. That work invoked  an outburst from the AGB star. Here we  see that model 1 and 2 conspicuously  show a bipolar outflow  even without the outburst. Repeating extended lobes are found both in the  north and south polar directions. The lobes can extend to $10-20$au with ripple-like structures caused by the pulsation. These  structures can also be affected by the binaries' orbital motion when we compare model 1 and model 2. In model 2, the secondary is more massive so its gravity has a more conspicuous imprint on polar outflows. In model 1 the ripples are more symmetric because  the secondary mass is only $0.1\text{M}_{\sun}$ and the dynamics of the polar outflow is  dominated by the radiation force from the AGB star.

\subsubsection{Circumbinary disc and accretion disc}

Circumbinary disc are common around AGB and RGB stars \citep{bujarrabal2013,kamath2016}. In our simulations, conspicuous circumbinary disc or tori have been found in models 1, 2 and 3. The circumbinary disk in model 2 resembles the simulation in \citet{mastrodemos1998}. The circumbinary disk in model 1 is optically thick in the equatorial plane and the material can fall back to the AGB star. The gap of the circumbinary disk in model 1 resembles a pioneer SPH simulation done by \citet{artymowicz1996}. The circumbinary disc in all the simulations are dynamically forming. They  expand outward and are replenished at their inner radii. Models 4 and 5 do not show circumbinary disc but instead exhibit spiral structure. 

Accretion disc are also found in models 1, 2 and 3. The x-plane cuts of the three models show that the accretion disc are stable in the equatorial plane and  are thin disc. The accretion disc do not exhibit 'flip-flop' instability in these three models. Possible reasons may be that the mass transfer mechanism in these models is not BH accretion \citep{matsuda1991,ruffert1996,bate1997a,bate1997b,foglizzo1999}, 3D simulation can suppress the instability \citep{soker1990,soker1991} and some physical process (i.e cooling, it somewhat comparable to making $\gamma$ closer to 1) is removing the excess energy thus may also suppress the instability. In all three models, the accretion disc are filling their Roche lobes and losing mass via the L2 point. The gas  leaking from the L2 point  merges into the circumbinary disc and can be pushed to higher orbit or escape. However, in models 4 and 5, no obvious accretion disc are found. 

\citet{huarte2013} carried out local simulations around the accreting star. finding  $\zeta\approx3$ per cent and accretion disc. Comparing our models 4 and 5 to their simulations, we identify three  reasons that conspire to prevent  accretion disk formation:
\begin{enumerate}
    \item Although we have  similar $\zeta$ values, our  AGB wind mass loss rate ($\dot{M}_\text{avg}=2.31\times10^{-7}$\msun\pyr)  is much smaller than  theirs ($10^{-5}$\msun\pyr).
    \item Our AGB wind velocity is higher and the mass of the secondary is smaller in our simulations than theirs. Both  result in less captured gas.
    \item The azimuthally averaged radiation force from the AGB star opposes  accretion disk formation.
\end{enumerate}

\subsubsection{Spiral structure}

Spiral structure around AGB stars has been discovered observationally \citep{maercker2012,mauron2006,kim2015} and  validated by simulations \citep{mastrodemos1998,edgar2008}. Figures in models 4 and 5 also successfully reproduce the spiral structure which is similar to CIT 6 and R Sculptoris. The spiral structure in our simulation is caused by BH accretion. In BH accretion, gas is focused behind the  secondary  due to under the gravitational force and cooled quickly due to the increased density. The gas conserves its outward momentum and angular momentum but loses its internal energy through cooling. The cold gas can keep its sharp density gradient for very long time. The  gas is focused into the equatorial plane where it  sweeps up more gas. Under some circumtances, L2 mass loss can also produce spiral structure  \citep{pejcha2016}.

The $x=0$  mid-plane cut of the spiral structure shows that gas  preferentially accumulates in the equatorial plane. The morphology of the $x=0$ plane cut is different from a torus or circumbinary disk. The  circumbinary disk and spiral likely represent two cases of a  'continuous' morphological sequence that depends on radiation pressure. To see this, first  compare the $z=0$ plane cut of models 4 and  5, we see that the curvature of the spiral structure in model 4 is greater than that of model 5. This spiral curvature  is influenced by the radiation pressure on the spirals; the larger the radiation pressure, the smaller the curvature. The radiation pressure is in turn, determined by the optical depth and   density in the equatorial plane. Thus the  equatorial plane density is  directly linked to the curvature of the spirals. Since the secondary in model 4, with its $8$au separation,  focuses more material into the equatorial plane than model 5 (with $10$au separation), the curvature of the model 4  spiral should be larger than that of model 5.

Furthermore, if the secondary focuses so much gas in its tail that the radiation force decreases more steeply with increasing radius than the gravitational force, the radius of the curvature of the spiral may become small enough (comparable to the binary separation), that the head and the tail of the spiral connect. The spiral structure will then become a circumbinary disk. Although we describe this transition between spiral and circumbinary disk as a  'continuous' sequence determined by the radiation pressure, the interaction between gas focused by the secondary, radiation force, and curvature of the spiral may make this transition range small and highly nonlinear. There remains opportunity to study this question further.

\subsection{The burst}\label{sec:burst}

The numerical setup of the burst simulation is represented by model 6 in table \ref{tab:modellist}. For this case, the AGB star incurs a burst from $t=30yr$ to $t=35yr$ (section \ref{sec:busrt}) and  maintains its  normal pulsating state before and after the burst. To demonstrate the distribution of the gas before and after the burst, we calculate average volume densities along $x$ and $z$ axes respectively. By computing the respective local surface densities and dividing by the domain range of the projected axis over the region of averaging. 

Mathematically, we have 
\begin{equation}
    \rho_{x}\left(y_{0},z_{0}\right)=\frac{\sum_{x=x_\text{min}}^{x=x_\text{max}}\rho\left(x,y_{0},z_{0}\right)V\left(x,y_{0},z_{0}\right)}{\left(x_\text{max}-x_\text{min}\right)S_{y_{0},z_{0}}}
\end{equation}
\begin{equation}
    \rho_{z}\left(x_{0},y_{0}\right)=\frac{\sum_{z=z_\text{min}}^{z=z_\text{max}}\rho\left(x_{0},y_{0},z\right)V\left(x_{0},y_{0},z\right)}{\left(z_\text{max}-z_\text{min}\right)S_{x_{0},y_{0}}}
\end{equation}
where $V\left(x,y_{0},z_{0}\right)$ and $V\left(x_{0},y_{0},z\right)$ are the volume of the cubes, $S_{y_{0},z_{0}}$ and $S_{x_{0},y_{0}}$ are the projected area of the columns. $\rho_{x}\left(y_{0},z_{0}\right)$ and $\rho_{z}\left(x_{0},y_{0}\right)$ ($\rho_{x}$ and $\rho_{z}$ hereafter) are the average density of x axis and z axis correspondingly.

\begin{figure*}
    \centering
    \includegraphics[width=1.0\columnwidth]{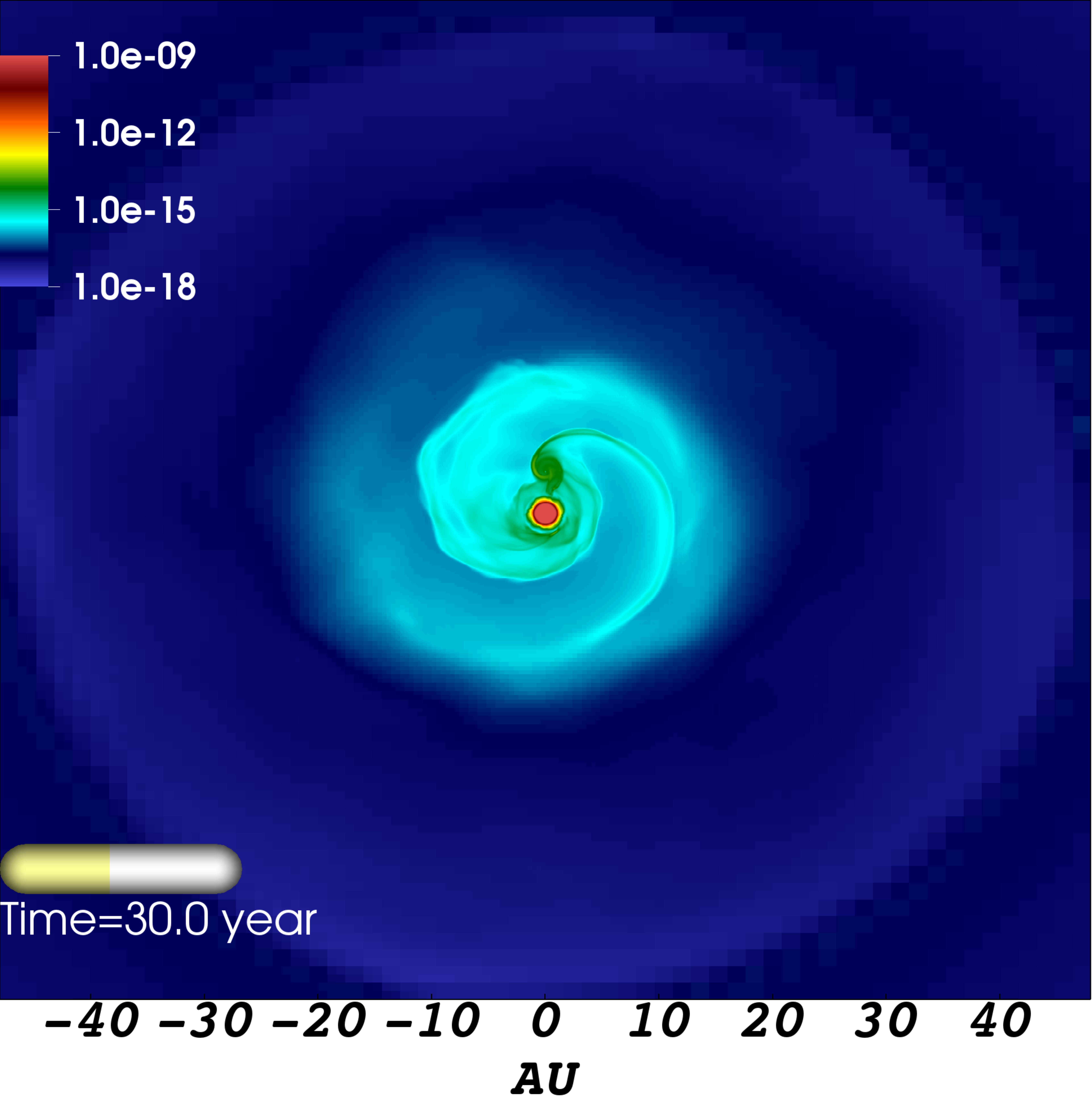}
    \includegraphics[width=1.0\columnwidth]{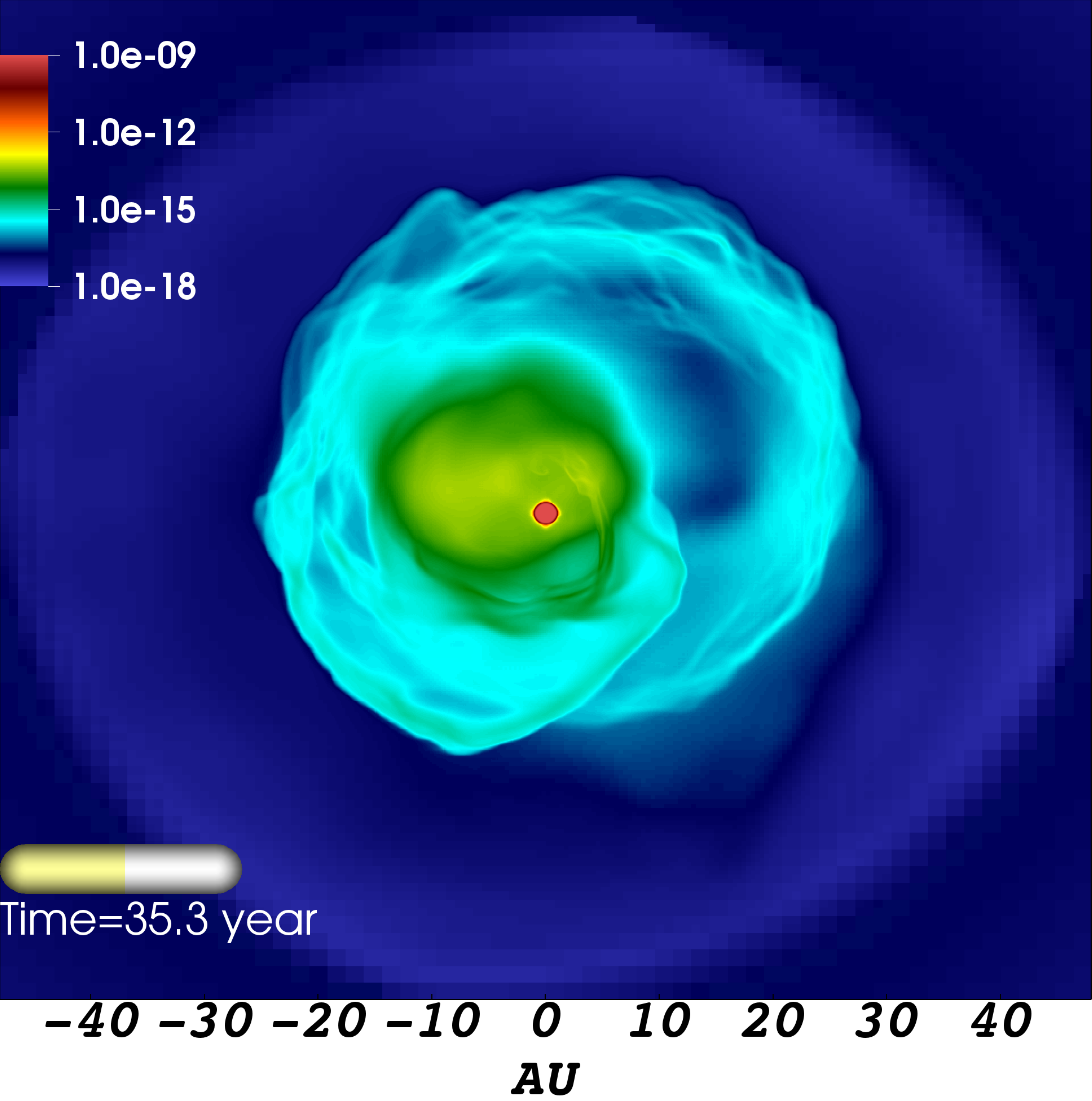}\\
    \includegraphics[width=1.0\columnwidth]{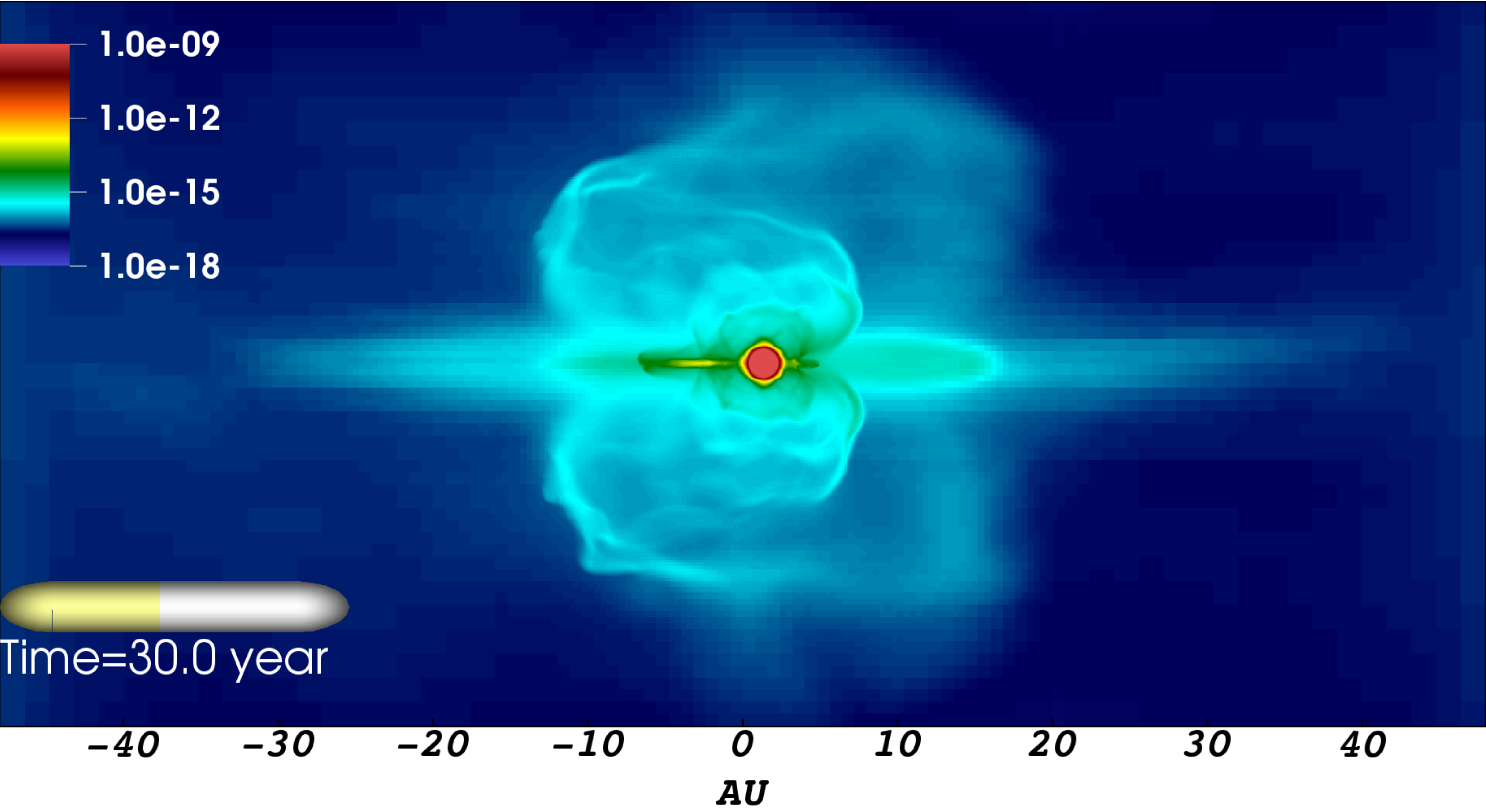}
    \includegraphics[width=1.0\columnwidth]{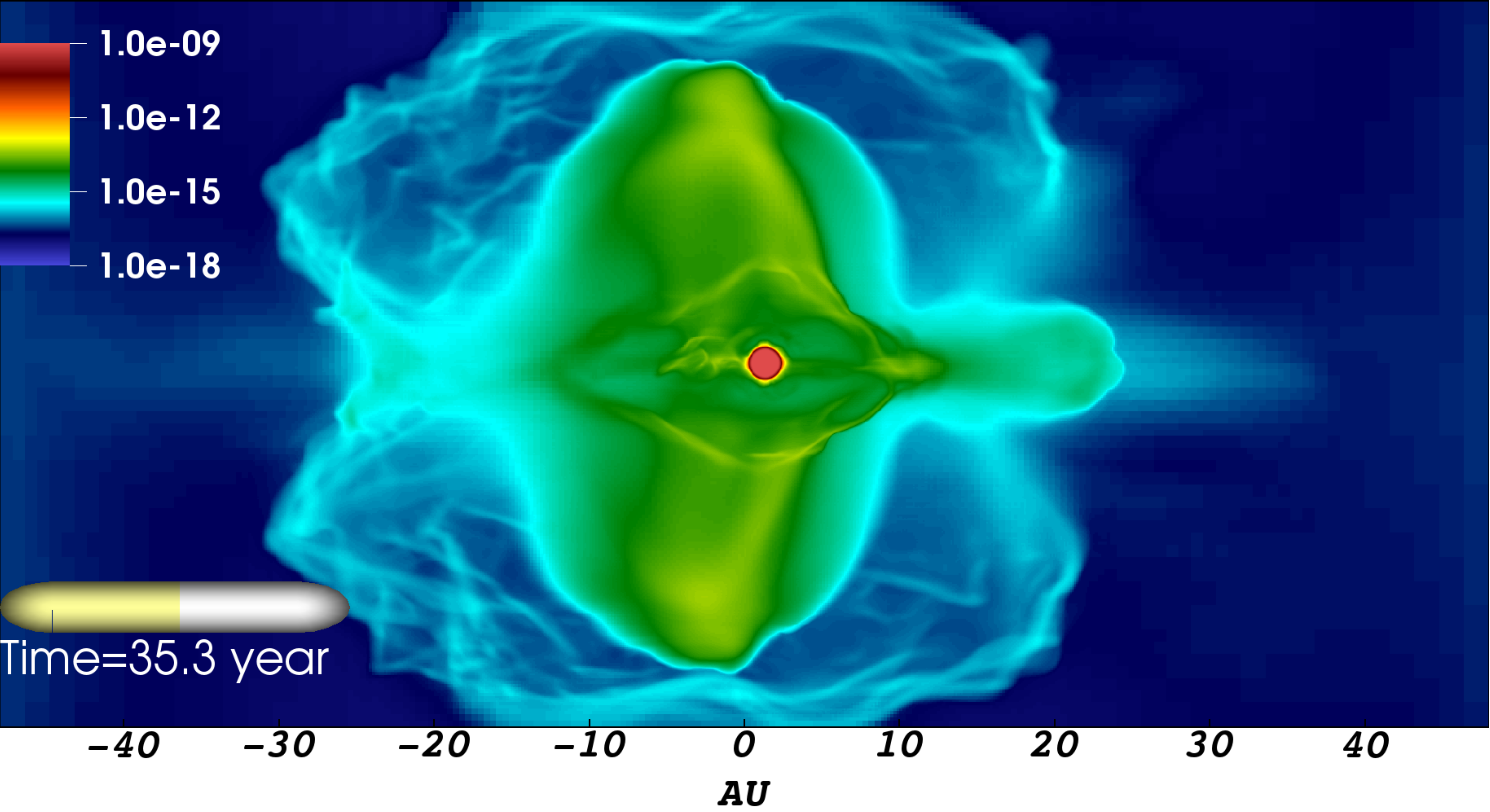}\\
    \caption{The odd rows display the average density of z axis projection in \gcc. The even rows display the average density of x axis projection in \gcc. The time stamps on the figures are increasing from left to right and from top to bottom.}
    \label{fig:burstbinarysim}
\end{figure*}

\begin{figure*}
    \contcaption{}
    \label{fig:continued}
    \includegraphics[width=1.0\columnwidth]{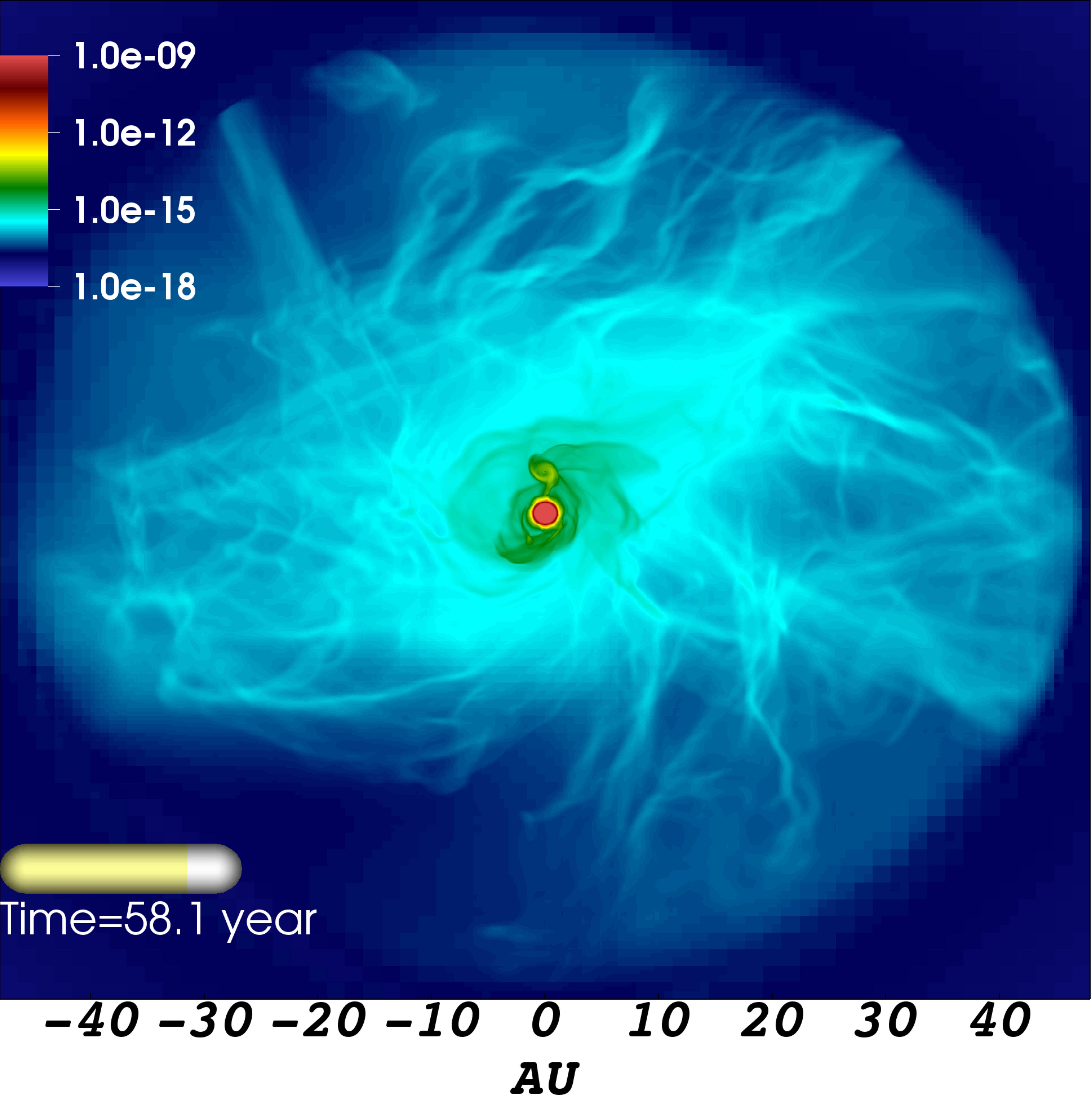}
    \includegraphics[width=1.0\columnwidth]{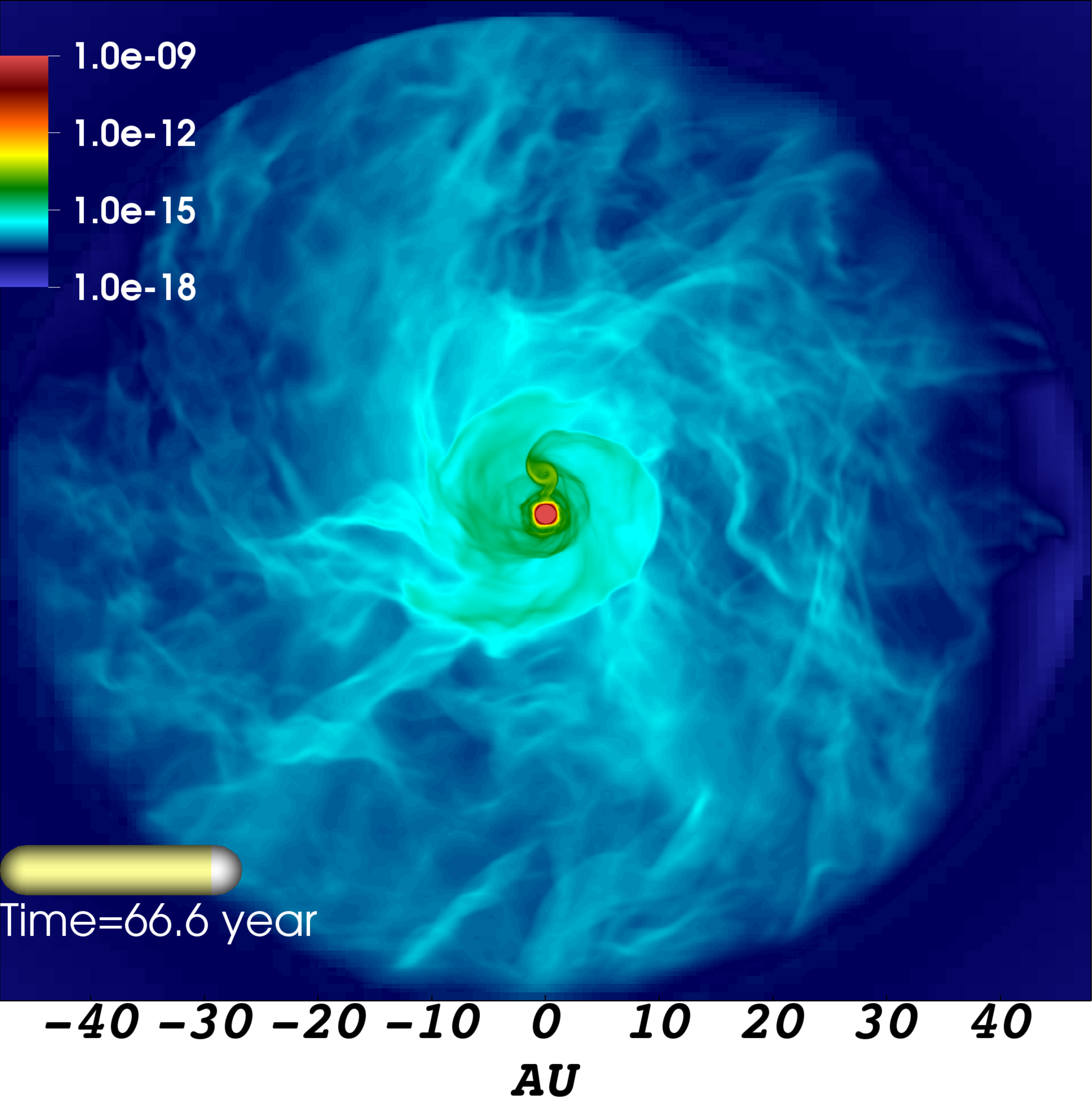}\\
    \includegraphics[width=1.0\columnwidth]{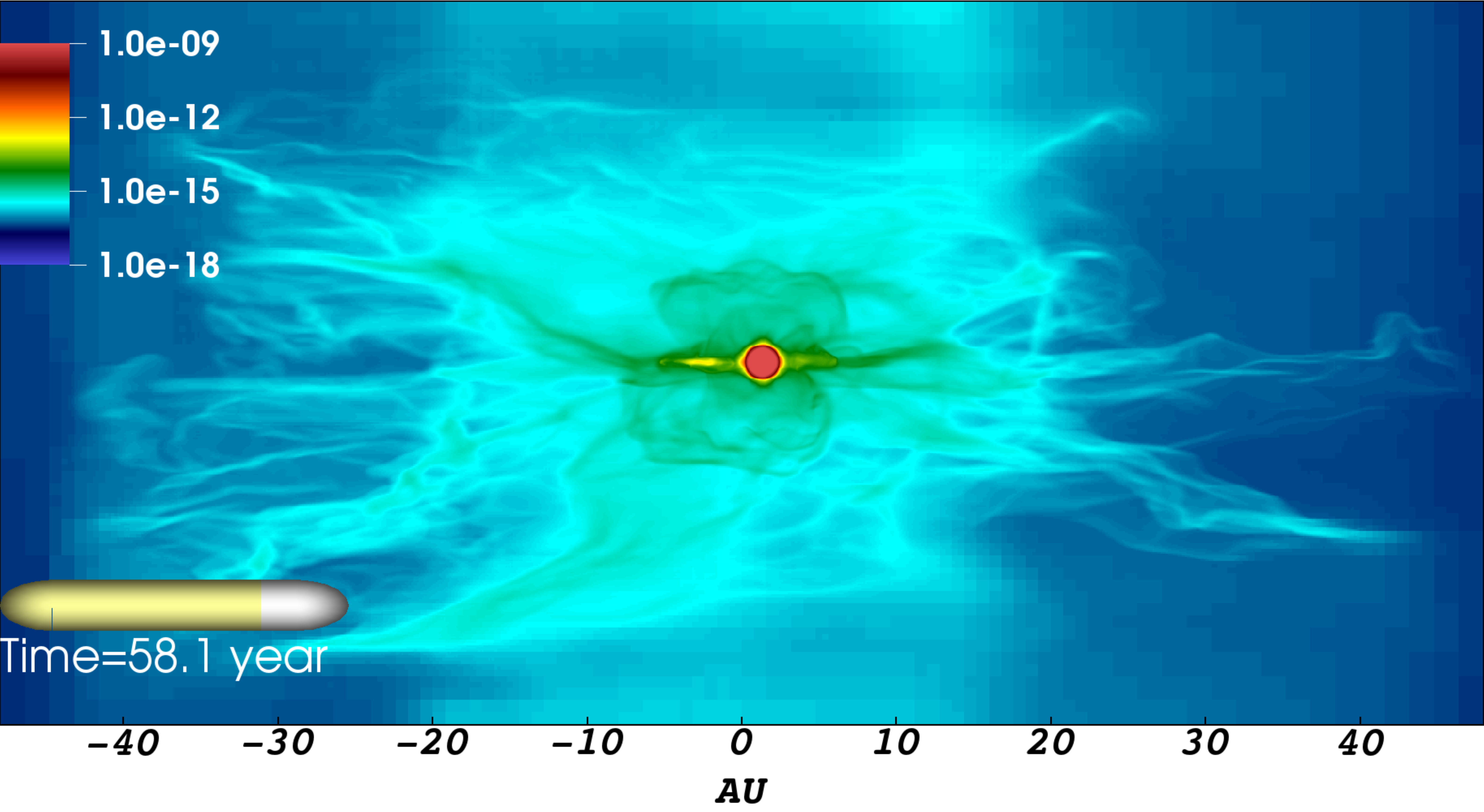}
    \includegraphics[width=1.0\columnwidth]{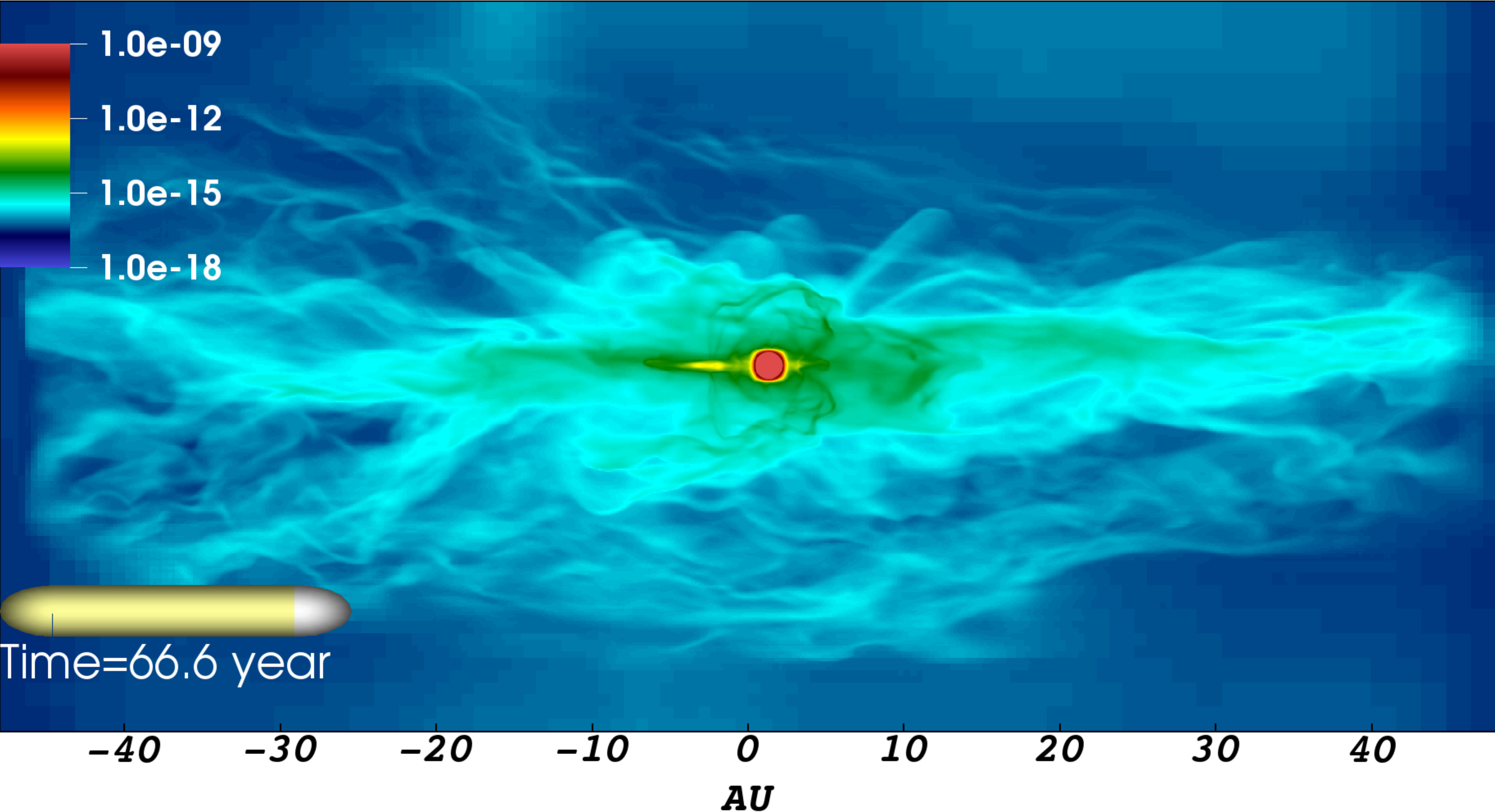}\\
\end{figure*}

Fig. \ref{fig:burstbinarysim} shows the the average density of $x$ and $z$  projections at different times. At $t=30.0$yr, the burst is just about to start and the binary  in the same state as model 2. We can therefore  compare this state with   the mid-plane-cut figures of model 2. The figure in the left column of the first row shows $\rho_{z}$, and reveals the accretion disk around the secondary and the circumbinary disk. There are two arms stretching out from the L2 and L3 points. The shape of the outer boundary is also seen. The figure in the left column of the second row shows $\rho_{x}$ in which the  accretion disk and the circumbinary disk are now seen edge-on. The figures also are also spiral arms in the polar direction which look very similar to those in L2 Puppis \citep{kervella2015,kervella2016}. 

The  first figure in the right column of Fig. \ref{fig:burstbinarysim} shows $\rho_{z}$ and $\rho_{x}$ at the later time $t=35.3$yr. At this time, the burst is over but we can see that the ejected material during the burst was asymmetrically distributed. In the equatorial plane, there is a ring-like structure and the accretion disk is  temporarily absent. The polar outflow  has a cigar-shaped structure. There is a low density 'hollow' region around the center in the $x=0$ plane cut residing inside the high density ejecta (see Appendix).  The ejected material will continue to expand while gas farther in is falling  toward the binary stars, resulting in the cavity. This gas can be accreted by the secondary, re-enter the AGB star's boundary and disappear from the simulation, or remain  in the equatorial plane and form a larger circumbinary disk. 

By $t=58.1$yr (third row of  Fig. \ref{fig:burstbinarysim})
the  gas in the polar direction  has been falling back for many years. The flow become turbulent and  shocks created by the colliding flows are evident throughout the region. The thin accretion disk around the secondary has fully formed. The  region very close to the binary star no longer resembles that of  model 2 as  there are no  arms  stretching arms from the L2 and L3 points. This region will continue to  evolve and eventually return to a stable state. 

At $t=67.9$yr, the accretion disk can be seen in both $x$ and $z$ axis projections. There is a circumbinary disk forming and we can see the two  arms again stretching from the L2 and L3 points. The center region is now returning to the stable state that resembles this region model 2. However,  much of the fall back material is orbiting the binary at larger radii, which is very different from the result of model 2. The polar gas  is still falling back on to the equatorial plane and colliding with the  AGB wind pulses. This create shocks which cool and lead to filament structure in the density projections. We believe that the simulation is still evolving but following the simulation further was too computational demanding for present purposes:
 at current resolution, to simulate $1$yr of this model cost 1400 CPU hour.

\begin{figure}
    \centering
    \includegraphics[width=1.0\columnwidth]{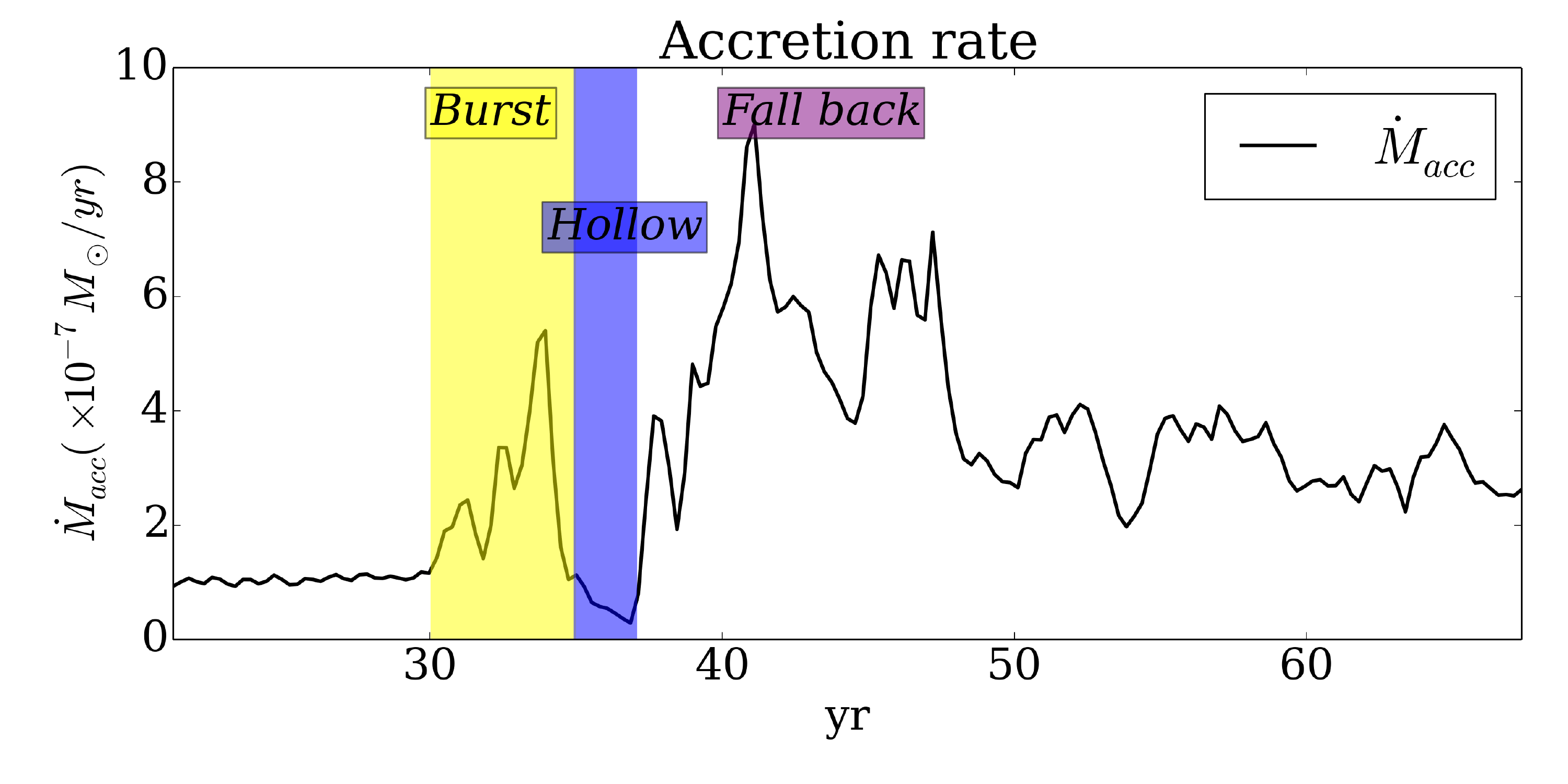}
    \caption{The accretion rate vs time.}
    \label{fig:accrete}
\end{figure}

\subsection{Accretion rate versus time}
 
We present the time dependent accretion rate of Model 6 in Fig. \ref{fig:accrete}.  Before the burst ($t=30.0$yr), the accretion rate is roughly $\dot{M}_\text{acc0}=1.1\times10^{-7}$\msun\pyr and   oscillates slightly with the AGB pulsation.  This value is the comparable to the accretion rate for  model 2 in table \ref{tab:masslosstable}. Since we use one more level of AMR in model 2 around the accreting star, the similar accretion rate implies that the resolution around the secondary is not influencing this quantity. 

During the burst, indicated by the yellow strip in  \ref{fig:accrete}, the accretion rate reaches
$5.0\times10^{-7}$\msun\pyr, or 5 times of the $\dot{M}_\text{acc}$. The ejected material during the burst  leaves a small 'hollow' region within. This causes the accretion rate to drop, reaching as low as  approximately half  of $\dot{M}_\text{acc0}$.

The final fall-back phase is perhaps the most interesting. Here the accretion rate can surge to more than $8.0\times10^{-7}$\msun\pyr or 8 times $\dot{M}_\text{acc0}$. As the material continues to fall back, the accretion rate decreases slowly, sustaining for more than $20$yr.

The two peaks and the prolonged decay of the accretion rate may lead to interesting time a of the binary light curves or temporal variability and spatial variability (i.e. knots) in jets.   \citep{lopez1993,soker1994,corradi1999,blackman2001,balick2002}.

\section{conclusions and discussion}\label{sec:summary}
To study mass transfer in binary systems with an AGB primary, we perform 3D radiation hydrodynamic simulations with {\small ASTROBEAR}. We carry out analysis from first principles and examine the physical properties of isolated AGB stars (section \ref{sec:isoagb}). Both normal pulsating states and bursting states are explained in detail and discussed \citep{vanloon2005,ivanova2016,staff2016b}. We perform six binary simulations with {\small ASTROBEAR}. Five of the binary models have a normal pulsating AGB star and one of the binary models has a bursting AGB star (Sec. \ref{sec:binary}).

Our isolated AGBs star produce AGB winds with reasonable terminal velocities and velocity profiles (Fig. \ref{fig:isolatedpulse}) but low mass-loss rates. We think it is mainly the incorrect EOS and lack of radiative-convective envelop that cause this small mass loss rate. In radiation hydrodynamic binary simulations, EOS and radiation pressure can induce convection and determine density profiles, which in turn sets the optical depth which effects the dynamics and mass-loss rate. Therefore, in principle, different EOS and opacity profile would yield different dynamics and mass loss rate. On the other hand, there is a similarity between high mass-loss rates in wide binaries and low-mass loss rates in close binaries as the density drops with distance. Therefore it is still worthwhile to carry out simulations in binary systems with low mass-loss rates. In particular, self-consistent AGB wind models are important and are of interest.

All binary models with the normal pulsating AGB star exemplify BH accretion and WRLOF. The ratio of accretion rate to mass-loss rate, $\zeta$, and the mass-loss rate enhancement, $\chi$, are calculated for each model. At least three mechanisms are responsible for the non-zero $\zeta$ and $\chi$ values in close binary stars (Sec \ref{sec:affmlrateandaccrate}). The results from model 1 to model 5 are displayed in Fig. \ref{fig:binarysimulation}. Models 1 and 2 exhibit clear evidence of bipolar outflows. Circumbinary disc and accretion disc can be found in models 1, 2 and 3. Spiral structure is observed in model 4 and model 5. We do not find any obvious accretion disc in model 4 and model 5. By comparing them to \citet{huarte2013}, we list the possible reasons why we do not see the accretion disc in our models. We successfully reproduce all three types of structures (bipolar, circumbinary disc and accretion disc and spiral structure) with the same mass-losing AGB star. We believe the versatility proves the broad correctness of our binary model.

The binary simulation with the bursting AGB star (Sec. \ref{sec:burst}) has the most intricate structure. We present both plots of the middle-plane cuts (Appendix) and projections. At first, it is similar to model 2. We do this with intent, as this demonstrates that the resolution is not adversely effecting the accretion rate and the dynamics. After the burst, the ejecta pushes the ambient material significantly outward, thereby creating a hollow inner region. The accretion rate drops during this time (Fig. \ref{fig:accrete}), afterwhich, we observe significant amounts of gas falling back. During this time, the infalling gas collides and shocks. The high density regions cool faster and conserve their filament structures. At the end of our simulation, we can see the actively forming circumbinary disk and accretion disk at the inner region and the evidence of the burst at large radii in the equatorial plane. The accretion rate shows two peaks, the first one corresponds to the burst enhancement and the second one corresponds to the fall back process. We can see prolonged high accretion rates due to the fall back process. This accretion rate change may provide clues on the formation of jets and knots in binary stars.

Lastly, we address two numerical difficulties in binary simulations.
\begin{enumerate}
    \item Two important structures in AGB binary systems are accretion disc and the AGB star itself. Radiation transfer is equally important in both objects. For now, we only resolve the radiation transfer around the AGB star with 2D ray-tracing method because of efficiency. Global 3D radiation transfer will be the best way to resolve the radiation transfer process in close binary systems but is currently limited due to computational complexity.
    \item In corotating simulations, we find that numerical viscous heating is a spurious, adverse effect. Numerical  heating is rooted in Eulerian codes as they suffer from the so called 'Galilean non-invariance' problem \citep{springel2010}. \citet{robertson2010} did a critically study on this problem and claimed that simulations done by Eulerian codes can converge to the reasonable solution if the resolution is high enough. In our simulations, we enlarge the resolved area and increased the level of AMR as the ejecta expand to keep the numerical viscosity low. We do this because the Mach number in the lab-frame at large radii is usually around $10$, but it may increase to $20$ in the corotating frame as the bulk velocity is Galilean variant. If the numerical viscous heating converts $1$ per cent of the kinetic energy to internal energy in $1$yr of simulation, the temperature will rise $4$ times in $1$yr. However, the increasing of temperature will be slowed down when the Mach number is lowered. One obvious effect is that the circumbinary disk may be puffed up and even disappear resulting in no Keplerian disc \citep{ruyter2006,bujarrabal2013} which should be observed at large radii. This effect is stronger in close binary simulations as they have higher angular frequency. Cartesian meshes may also worsen the situation. To better understand dynamics of circumbinary disc, we think both Lagrangian fluid codes or Eulerian codes with cylindrical or spherical meshes would be beneficial.
\end{enumerate}

\section*{Acknowledgements}

We gratefully acknowledege all the staffs in Center for Integrated Research Computin (CIRC) for supporting us with the amazing computer resources, we can not finish the simulations so fast without their help. Special thanks should be given to Dr. Baowei Liu for helping us setting up the code. We also thank Mr. Richard E. Sarkis and Mr. David Munson for technical support. We appreciate the discussion with Prof. Bruce Balick and Prof. Hans van Winckel on AGB binary problems and Prof. James Stone on the 'flip-flop' instability.




\bibliographystyle{mnras}
\bibliography{mnras} 



\appendix
\section{the 'Hollow' in model 6}

\begin{figure}
    \centering
    \includegraphics[width=0.49\columnwidth]{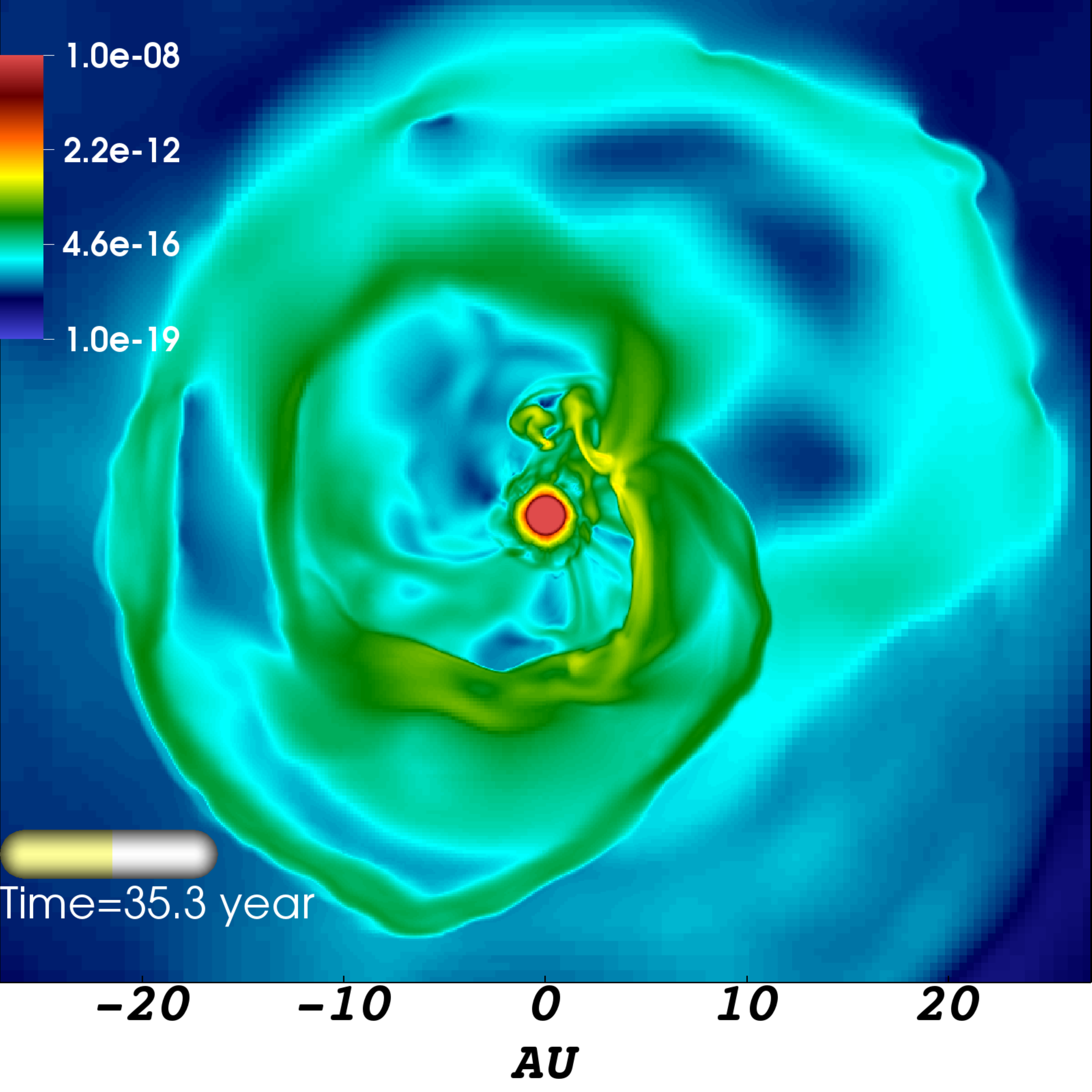}
    \includegraphics[width=0.49\columnwidth]{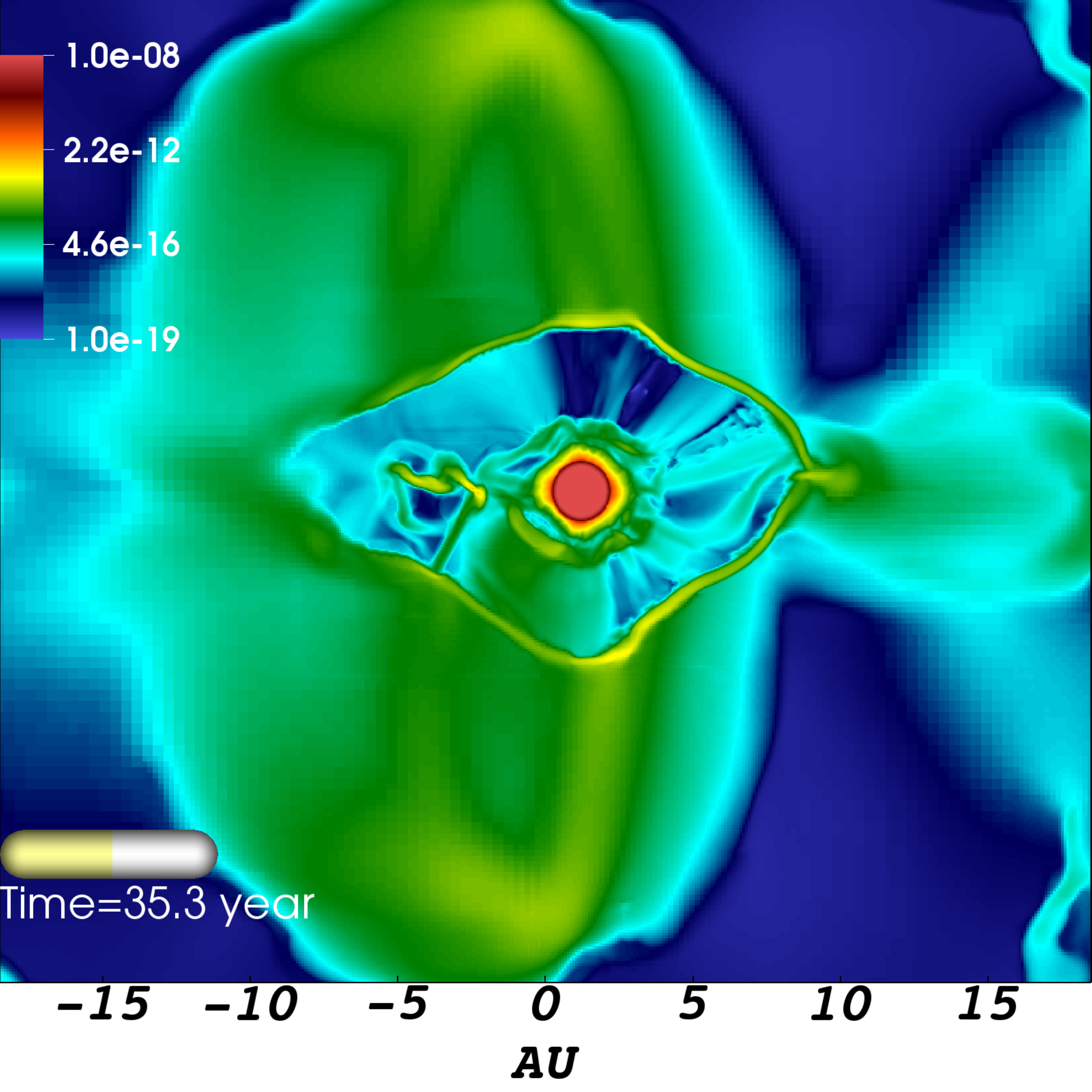}\\
    \includegraphics[width=0.49\columnwidth]{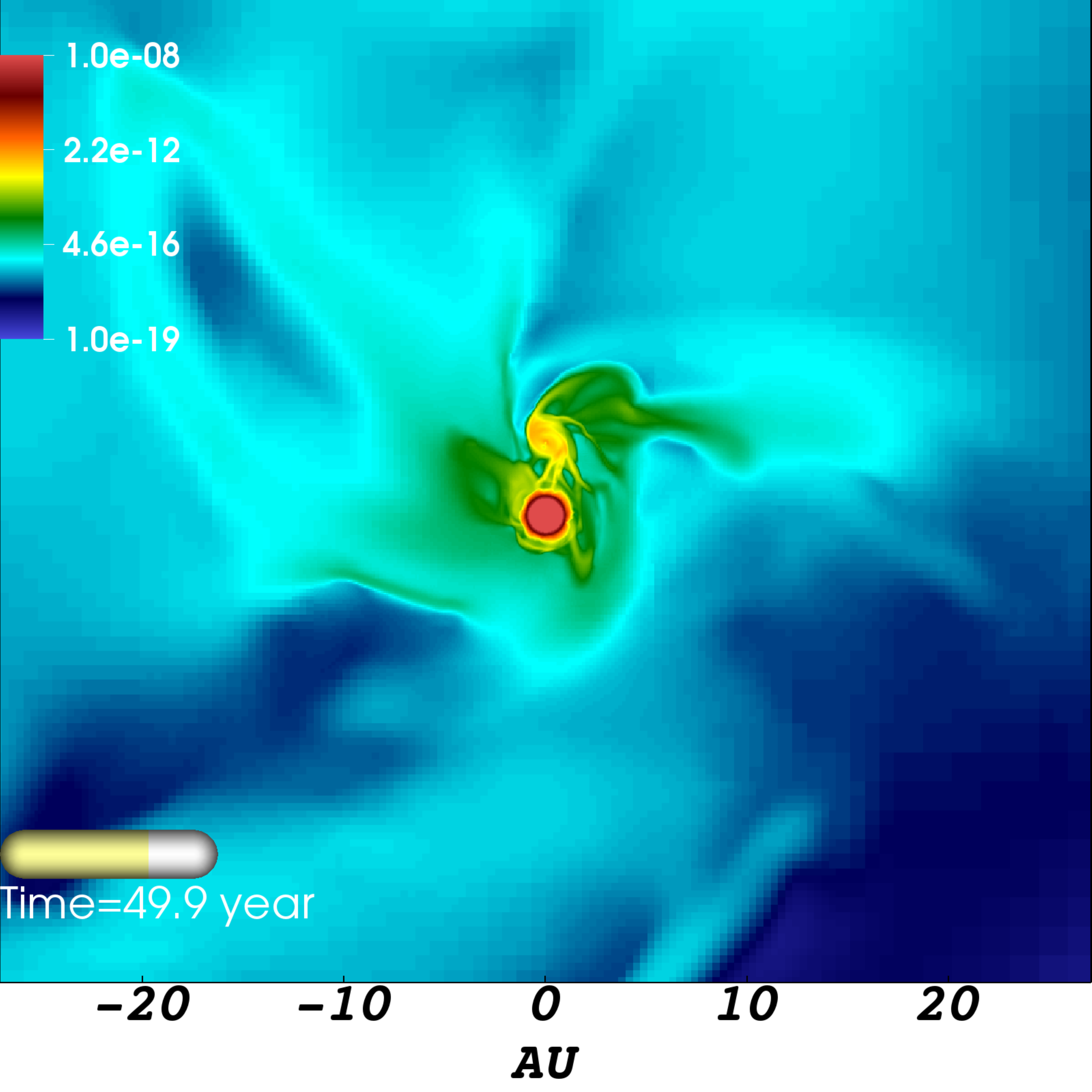}
    \includegraphics[width=0.49\columnwidth]{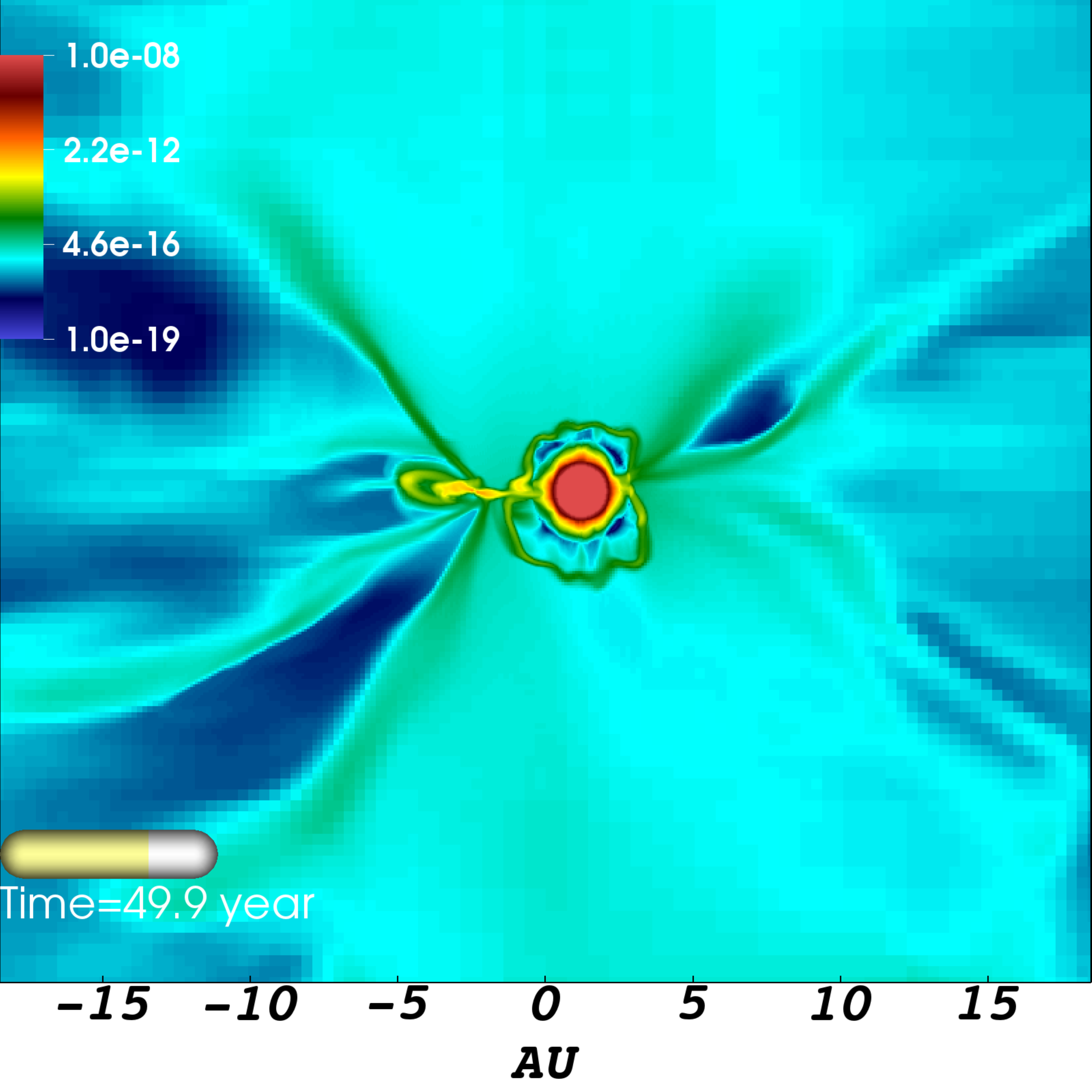}\\
    \includegraphics[width=0.49\columnwidth]{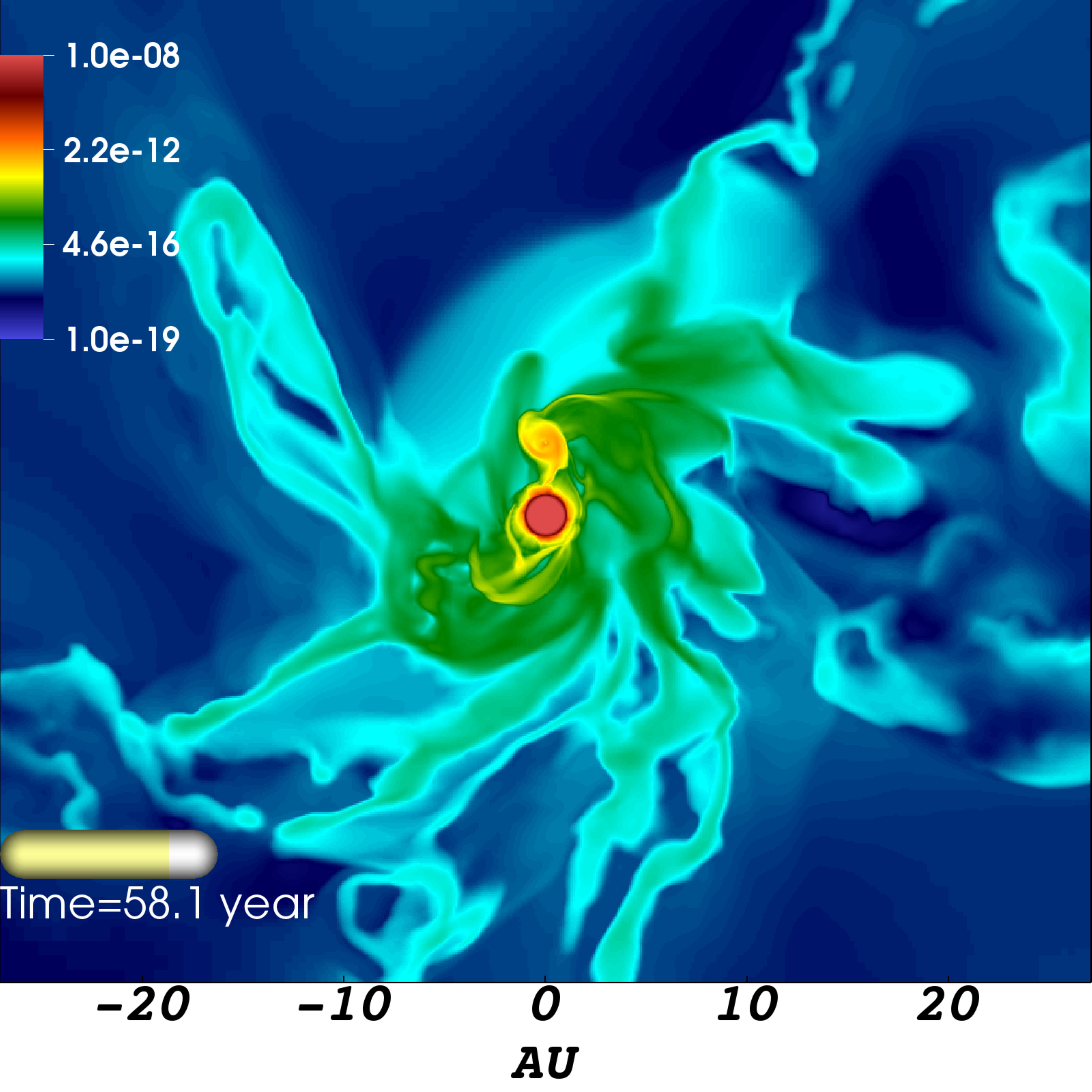}
    \includegraphics[width=0.49\columnwidth]{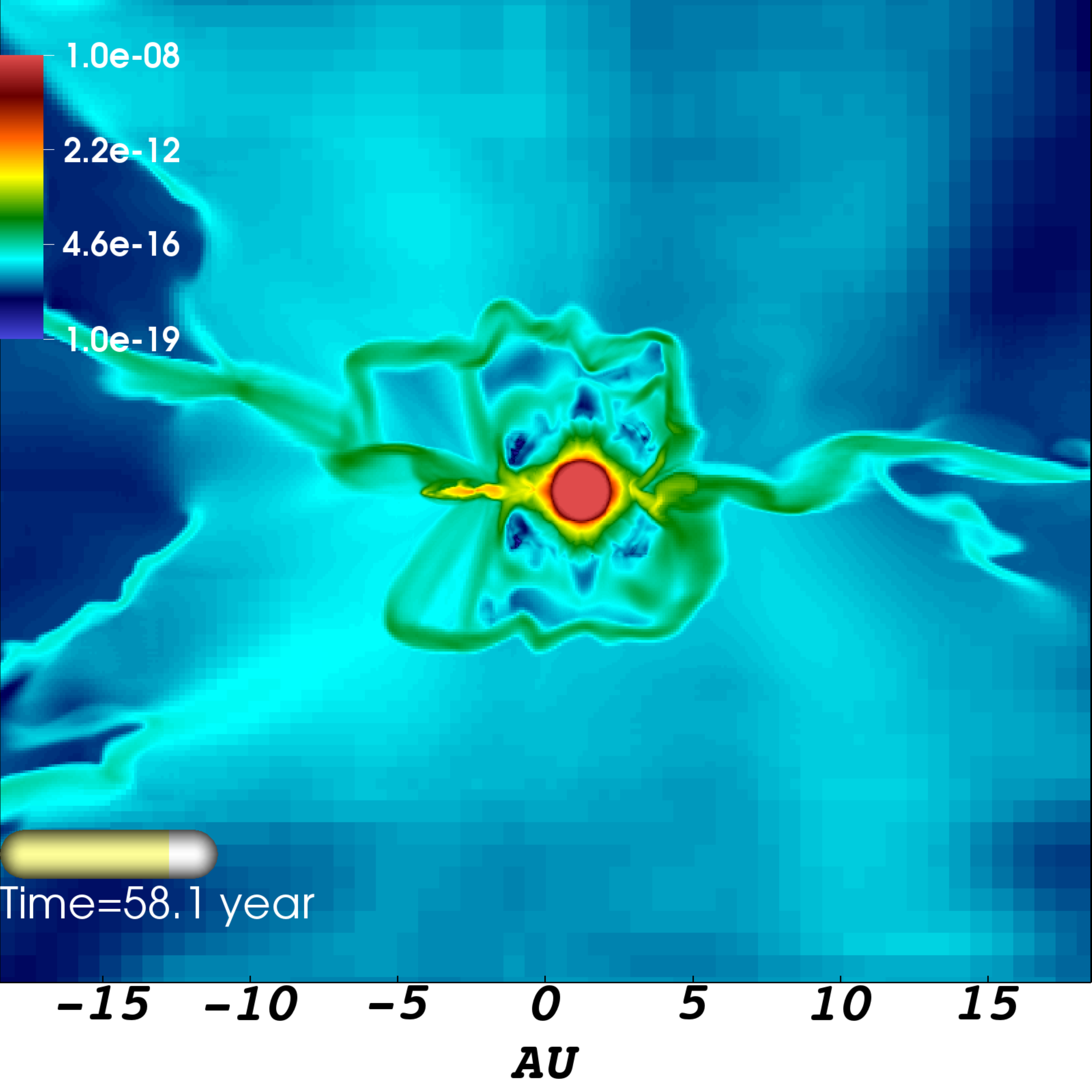}\\
    \caption{From top to bottom are the density plot in $g/cm^3$ at different time in model 6. The figures on the left are the $z=0$ plane cut and the figures on the right are the  $x=0$ plane cut.}
    \label{fig:hollow}
\end{figure}

Figure \ref{fig:hollow} shows the middle plane cut plots from $t=35.3$yr to $t=58.1$yr. During this time, the gas is falling back and it is more clear to see the structure in middle plane cut plots than the projections.

We can clearly see the 'eye shape' low density region inside the ejecta at $t=35.3$yr. It is 'closing' with the fall back of gas. The fall back gas will collide into each other and create shocks. It can also interact with the successive AGB wind as can be seen at $t=58.1$yr.


\bsp	
\label{lastpage}
\end{document}